\documentclass{svjour2} 

\usepackage{hyperref,graphicx,url,amssymb}

\smartqed  
\begin{document}
\title{Modeling Translation in Protein Synthesis with TASEP: a Tutorial 
and Recent Developments}

\titlerunning{Modeling Protein Synthesis with TASEP}

\author{R.K.P.~Zia, J.J.~Dong, and B.~Schmittmann}

\institute{R. K. P. Zia \at
         Physics Department, Virginia Tech\\ Blacksburg, VA, 24061, USA \\
              \email{rkpzia@vt.edu}
           \and
           J. J. Dong \at
               Physics Department, Virginia Tech\\ Blacksburg, VA, 24061, USA \\
             Physics Department, Hamline University\\St. Paul, MN, 55104, USA\\ 
             \email{jdong01@hamline.edu}
           \and
           B. Schmittmann \at
         Physics Department, Virginia Tech\\ Blacksburg, VA, 24061, USA \\
              \email{schmittm@vt.edu}
          }

\date{Received: date / Accepted: date}

\maketitle

\begin{abstract}
The phenomenon of protein synthesis has been modeled in terms of totally asymmetric simple exclusion processes (TASEP) since 1968. In this article, we provide a tutorial of the biological and mathematical aspects of this approach. We also summarize several new results, concerned with limited resources in the cell and simple estimates for the current (protein production rate) of a TASEP with inhomogeneous hopping rates, reflecting the characteristics of real genes. 
% \PACS{05.70.Ln Nonequilibrium and irreversible thermodynamics 
%\and
%87.10.Mn Stochastic modeling 
%\and 
%87.15.hj Transport dynamics}
\end{abstract}

\section{Introduction}

Nonequilibrium statistical physics remains one of the greatest unsolved
challenges of theoretical physics. In recent studies, both the National
Academy of Sciences \cite{NAS} and the Department of Energy \cite{DOE} have
recognized the importance and scientific impact of developing a fundamental
and comprehensive understanding of physics far from equilibrium. Unlike
string theory or cosmology, this field addresses phenomena in our immediate
experience, such as the flocking of birds or fish  
\cite{Feder07,SKPF03,BCCCCOPPVZ08}, traffic flow \cite{CSS99,CSS00,PSSS01}, 
or biological transport \cite{Howard01,SW03,KF07,Friedman08,MGP68,MG69}, 
to mention
just a few examples. Living systems and biological phenomena, in particular,
are areas where concepts and methods from nonequilibrium statistical physics
find a natural application. On the one hand, nonequilibrium statistical
physics is concerned with open many-particle systems, sustaining nontrivial
currents of energy or particles; and on the other hand, biological systems
are characterized by considerable complexity and depend on energy and matter
throughputs for proper functionality. Of specific interest are biological
transport phenomena restricted to an effectively one-dimensional track, such
as kinesin and dynein on microtubules, RNA polymerase on DNA during
transcription, and ribosomes on mRNA since these offer the promise of being
accessible via simple model systems. The last process, usually referred to
as translation, is a key element of protein synthesis, and will be our prime
focus in the following.

One possible approach is to start with a good understanding of the physical
and biological processes involved in protein synthesis, so that a detailed
model can be developed and tailored to a particular experimental situation.
While this promises the possibility of quantitative comparisons between
model results and experimental data, any underlying generic characteristics
cannot be easily identified, due to the large amount of experimental detail
involved. An alternate approach - which has proven immensely powerful in
equilibrium statistical mechanics - is to study simple model systems which
can provide deep insights into generic behaviors and universal phenomena.
Even though such simple models may not allow for immediate comparisons with
experimental data, due to the considerable amount of simplification or
abstraction involved, they can still guide future experimental work,
especially when good data are not yet available. In summary, both approaches
have their strengths and drawbacks and offer complementary insights.

In this article, we discuss one particular marriage between statistical
physics and biology. We will approach a very complex biological process -
protein synthesis - starting from a simple model, the totally asymmetric
exclusion process (TASEP). In a twist of history, this model was first
suggested in this context in the late 1960s \cite{MGP68,MG69} and endowed with
as much experimental detail as known at the time. Independently, models of
stochastic directed transport were proposed and studied by mathematical
physicists \cite{Spitzer70}. About two decades ago, these were (re-)
discovered by the nonequilibrium statistical physics community \cite{Krug91}
and became paradigms of this field, much like the Ising model for the study
of phase transitions and critical phenomena.

Our goal here is two-fold, namely, to provide a brief tutorial and to
discuss some recent developments. The next two sections are devoted to the
former and the following one to the latter. 
Thus, Section \ref{Rudi} is designed for
those new to biology, providing some basics of how proteins are synthesized
in a cell and, more specifically, a part of the process known as
``translation.'' 
Complimentarily, Section  \ref{TASEP} is designed for readers who
may be interested in getting involved with exploiting TASEP as a model for
translation. Assuming they are familiar with the basics of statistical
mechanics and simulation techniques, this section is meant to provide a
self-contained, though by no means comprehensive, review of the simplest
TASEP \cite{Spitzer70} and generalizations relevant for translation. Section
IV is devoted to two recent developments. The first is a brief review of
competition between multiple TASEPs. In the second, we are motivated by
``silent mutations'' of genes to introduce a novel notion in statistical
mechanics, namely, a quenched distribution of distributions. We end with a
brief summary and outlook in Section \ref{sum}.

\section{\label{Rudi}Rudiments of protein synthesis}
 
Essentially every vital process within cells of living organisms involves
proteins (a.k.a. polypeptides). These are macromolecules formed from chains
of amino acids. For example, hemoglobin is a protein which carries oxygen
from our lungs to the rest of our body. The blueprint for how these proteins
are synthesized is contained in the DNA of the cell. Over the past century,
biochemists gradually discovered the way this information is transformed
into a physical molecule. Much of this body of knowledge is quite well
established and can be found in many standard texts of microbiology \cite
{cell}. For the reader's convenience, this Section provides a brief synopsis
of this process.

Protein synthesis involves two stages: {\em transcription} of genetic
information from DNA to messenger RNA (mRNA) by RNA polymerase and {\em %
translation} from mRNA to proteins through ribosome translocation.
Articulated by F. Crick in \cite{Crick58,Crick70}, this central dogma of
molecular biology captures the essence of transferring sequence information
to functional macromolecules (e.g. RNAs and proteins) in all life forms. One
of the most complex cellular processes, protein synthesis demands concerted
actions by hundreds of molecules in sequential steps and typically requires
a high level of regulation. Its vast demand for the energy needed to
complete the reactions also establishes its crucial role in all metabolic
pathways. Therefore, developing a {\em quantitative understanding} of
transcription and translation processes would be most desirable. Indeed,
this task has dominated much of recent research in molecular biology, as
well as mathematics, physics and emerging cross-disciplinary fields. Of
course, any model that encompasses all the biochemical reactions and the
structural components in translation will be prohibitively complex. Rather,
it is more practical to attempt at gaining some insight into the process of
transforming DNA information into a polypeptide chain by restricting our
attention to single-cell organisms and identifying the most essential
ingredients. To this end, we focus on the bacterium {\em E.coli}, a
well-established model organism in molecular biology of which abundant
genetics and kinematics data are available for further analysis. In this
article, we will further restrict our attention to the second stage:
translation. Our aim here is to condense the relevant biological information
into a simple overview of protein synthesis, so that interested readers may
actively participate in model-building/analysis.

In most bacteria such as {\em E.coli}, translation involves three main
players: the mRNA (genetic template), the ribosome (assembly machinery), and
aminoacyl transfer RNAs (aa-tRNAs), i.e., transfer RNAs ``charged'' with the
corresponding amino acid.

The mRNA carries genetic information, encoded as triplets of nucleotides.
Each triplet is known as a ``codon''. Since there are four nucleotides (A,
U, C, G), there are $4^{3}=64$ distinct codons, e.g., AUG, CGG, etc. Except
for three ``stop codons'' (UAA, UGA, and UAG) that signal the termination of
translation, each of the remaining 61 ``codes'' for one of the 20 amino
acids. Each of the latter is conveniently denoted by a single letter: A, C,
D, E, F, G, H, I, K, L, M, N, P, Q, R, S, T, V, W, and Y. Indeed, we may
regard a protein as a word, i.e., a string of letters from the above list,
ranging from $\thicksim 10$ to $\thicksim 1000$ long. With 61 codons and 20
amino acids, there can be, {\em on the average}, three codons associated
with the same amino acid. As a result, a protein with $L$ amino acids can be
coded by $\thicksim 3^{L}$ distinct sequences (strings of codons or mRNAs).
To complicate matters, some amino acids are coded by a single codon (e.g.
AUG for methionine/M and UGG for tryptophan/W) while others are associated
with as many as 6 (e.g. CGU, CGC, CGA, CGG, AGA, AGG for arginine/R). Therefore, the exact
protein-mRNA degeneracy depends on the sequence (though not the precise
order of codons). Codons coding for the same amino acid are termed
``synonymous.'' In the example above, CGU and AGG are synonymous codons. Two
sequences which differ only by synonymous codons are known as ``silent
mutations,'' in the sense that both produce the same protein (polypeptide
chain). In addition to the string of codons, there is a long sequence of
nucleotides at the beginning of the mRNA, known as the ``Shine-Dalgarno
sequence'' (SD sequence) \cite{SD75}. This region controls the binding of a
ribosome to the mRNA, which is also the start of the process of translation.
Known as ``initiation'' in biology, even this starting event is quite
complex \cite{Kozak83,Moldave85,NN90,Merrick92,PH00,LSMS05,APLE06}, 
requiring the presence of several initiator proteins.

The next major player is the ribosome, a sizable molecule composed of a
large and a small subunit. Within the ribosome, there are three sites to
which a tRNA can bind and unbind. Designated as A, P, and E, these are,
respectively, the aminoacyl site (for docking of an aa-tRNA), the peptidyl
site (for transferring and binding the newly arrived amino acid to the 
partially formed polypeptide chain) and the exit site 
(for releasing the tRNA). 

The process
of translation consists of ribosomes moving along the mRNA without
backtracking (from one end to the other, technically known as the 5' end to
the 3' end) and is conceptually divided into three major stages: initiation,
elongation and termination. Among the three steps, initiation is of the
highest complexity and has seen significant developments in unravelling its
molecular details \cite{Kozak83,Moldave85,NN90,Merrick92,PH00,LSMS05,APLE06}
. Here, the ribosome interacts with the SD sequence through complementary
base-pairing and locates the start codon AUG with the help of several
initiator proteins. Translation begins with the assembly of the two
subunits of the ribosome, along with a tRNA charged with the M amino acid 
in the A-site.
The next steps, elongation and termination, are also quite intricate
\cite{RRCM90,SP91,BSWN94,CM96,NE98,FH05,NI98,KF99}. The ribosome moves 
along the mRNA, ``reading'' codon by codon, recruiting
the appropriate aa-tRNA, ``knitting'' the latest amino acid into the
partially completed chain, and releasing the ``discharged'' tRNA. This
cyclic process consists of the following steps. (i) The last amino acid of a
partial chain is attached to its tRNA at the P-site, aligned with a certain
codon. (ii) An aa-tRNA, correctly matched with the next codon, docks at the
A-site. (iii) The peptide bond between the amino acid and the tRNA at the
P-site breaks and reattaches to the new amino acid at the A-site. (iv) The
ribosome moves forward so that the two tRNAs are now at the E- and P-sites.
(v) The discharged tRNA in the E-site is released, leaving the A-site empty
for the next aa-tRNA. Finally, when the ribosome encounters one of the three
stop codons, the termination process commences: The ribosome disassociates,
while the completed amino acid chain is released (and folds into a
functioning protein). The whole process is quite involved and, instead of
providing a figure here, we direct interested readers to one of the many
helpful animations on the WWW \cite{proteinsyn}. 

Obviously, the third major set of players are the tRNAs. One end of this
class of molecules consists of one of the many anticodons (e.g., UAC, 
to match with AUG)\footnote{
Simplistically, there would be 61 anticodons to match the 61 codons.
However, nature is more complicated. Most cells contain less, due to 
``wobbling.'' For {\em E.coli}, there are 46 distinct anti-codons. We 
will ignore this extra complication here and discuss translation
as if there were 61 anticodons.}.  
The other end is an acceptor stem, to which an appropriate amino
acid (one of 20) can be attached, forming a ``charged'' aa-tRNA. Normally,
the mapping from amino acids to anticodons is one-to-many; details may be
found in, e.g., \cite{genecode}.
For our purposes, the main concern is the rather dissimilar set of{\em \
degeneracies}, i.e., the number of synonymous codons, $m_{aa}$, for the
amino acid $aa$: 
\begin{eqnarray}
&&\,\,
\begin{tabular}[b]{|c|c|c|c|c|c|c|c|c|c|c|}
\hline
$aa$ & A & C & D & E & F & G & H & I & K & L \\ \hline
$m_{aa}$ & 4 & 2 & 2 & 2 & 2 & 4 & 2 & 3 & 2 & 6 \\ \hline
\end{tabular}
\nonumber  \\
&&
\begin{tabular}[b]{|c|c|c|c|c|c|c|c|c|c|c|}
\hline
$aa$ & M & N & P & Q & R & S & T & V & W & Y \\ \hline
$m_{aa}$ & 1 & 2 & 4 & 2 & 6 & 6 & 4 & 4 & 1 & 2 \\ \hline
\end{tabular}
\label{aa-deg}
\end{eqnarray}
Note that 3 codons are reserved for termination, so the total here is only
61. Meanwhile, the concentrations of these aa-tRNAs in a typical cell are
known to be far from uniform. Indeed, for {\em E.coli}, the relative
abundance can be as much as a factor of 15  \cite{DNK96,VBL84,Ikemura81}. 
Since the elongation
rate is believed to be correlated with the aa-tRNA availability (as a
ribosome must ``wait'' for the appropriate aa-tRNA to arrive before
proceeding)  
\cite{DNK96,Maaloe79,SP91,KMTP09}, the time it takes a ribosome to complete
translation can vary widely, depending on the codon sequence. Further, the
rate of protein synthesis depends not only on the speed of a single
ribosome, but also on how many ribosome may be translating simultaneously
(on the same mRNA). Therefore, the wide range of degeneracies shown in (\ref
{aa-deg}) implies that silent mutations may have serious implications for
protein production rates.

Finally, for a cell to function properly, the important quantities to be
controlled are, presumably, the levels of various proteins. In a steady
state, a protein's level depends on the rate of its degradation as well as
its production. Assuming the former is the same for all proteins, then
the concentration of any particular type follows its production rate
closely, and the latter is just the average current associated with
translating that mRNA. Thus, our main interest here is the following
question: How are protein production rates correlated with specific sequence
information? Of course, translation in real biological systems is much more
complex, involving subtleties such as prokaryotes vs. eukaryotes, initiation
and elongation factors, wobbling, signaling, and regulation in response to
external conditions, etc. All are beyond the scope of this article, as we
focus on a few key ingredients, build the simplest models, explore their
behavior, and make some inroads into the remarkable processes of life.

\section{\label{TASEP}The TASEP and its generalizations}

In this section, we turn to another topic: the totally asymmetric simple
exclusion process (TASEP), which has been exploited to model protein
synthesis for over 40 years. After a brief historic introduction, we review
recent progress on simple models and generalizations to account for more
realistic conditions in biological systems.

\subsection{\label{history}Early history: Studies by Gibbs, {\em et.al.} and Spitzer}

Around 1970, TASEP was introduced from two entirely distinct motivations,
one from biology and the other from mathematics. Not surprisingly, the two
approaches are also quite different. In this subsection, we devote brief
paragraphs to each.

Gibbs and collaborators \cite{MG69,MGP68} were interested in a quantitative
description for translation, the process reviewed briefly above. The mRNA,
the codons, and the ribosomes are modeled by, respectively, a
one-dimensional open lattice, its sites, and particles (which enter the
lattice at one end, hop unidirectionally from site to site, and exit at the
other end). Aware of the large size of ribosomes compared to codons, these
workers began with particles which ``cover'' $\ell \geq 1$ sites. Further,
their formulation allowed for the possibility that particle hopping rates
can be bidirectional, as well as being site- and time-dependent. Setting up
the full stochastic problem, with a master equation for the probability to
find particles at each site, is relatively easy. However, no one has been
able, so far, to solve such a complex problem, even for the steady-state
distribution with time-independent rates. Nevertheless, there has been
significant progress since 1968, on various fronts. Historically, Gibbs, et.
al. focused on a system with uniform and time-independent rates, ignored
some correlations between ribosomes yet accounting for the strict exclusion
due to $\ell >1$, and set up recursion relations for $\rho _{i}$, the
density of particles at site $i$ (i.e., the average occupation of site $i$
by, say, the trailing edge of a particle), with the current $J$ as a
to-be-determined unknown. Their analysis of these recursion relations, using
both analytic and numerical techniques, led them to several important
conclusions. Examples include a non-trivial $\ell $-dependent relationship
between $J$ and the overall density $\bar{\rho}$, as well as the existence
of different phases as functions of the entry/exit rates. Since their model
is too far from real biological systems and experimental techniques in
biology were too primitive to probe occupations at the molecular level,
their results have lain largely dormant until recent years.

At the other front, Spitzer was interested in Markov processes involving
interacting particles \cite{Spitzer70}. The simplest interaction would be
just exclusion, i.e., each site can be occupied by at most one particle --
thus the ``simple exclusion'' part of TASEP. The simplest system would be a
one-dimensional periodic lattice, i.e., a ring of $L$ sites. Placing $N$
particles on such a ring and letting them hop randomly to nearest-neighbor
sites obviously leads to ``simple diffusion.'' Even in such a minimal
system, highly non-trivial behavior can be found \cite{DG09}, but
we will focus on a dynamics more closely resemble that of protein synthesis,
namely, hopping is allowed only in one direction. Thus, we turn to the
``totally asymmetric'' part of TASEP. Though the steady state is still
simple, its dynamics displays even more intricate properties
\cite{dMF85,KvB85,MB91,GS92,DEP93,Kim95,GM04,GM05}.
Of course, Spitzer also considered lattices with open boundaries and
particles hopping on/off the lattice at the ends. But this problem was
sufficiently more complex than the ring that even its steady state
distribution remained elusive for another two decades. In the next
subsection, we specify Spitzer's model and briefly summarize some of the
progress since the early 90's.

\subsection{\label{proto}The proto model and its properties}

\begin{figure}[tbp]
\begin{center}
\includegraphics[height=3cm,width=10cm]{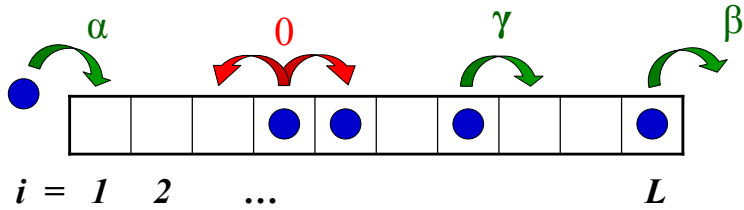}
\hspace{6cm}
\caption{Schematic summary of the
rules of the proto TASEP. Particles cannot hop backwards or onto an occupied
site (red online). They move forwards to empty sites with rate $\gamma $
(green online). Entry and exit rates are denoted by $\alpha $ and $\beta $,
respectively (green online).}
\label{fig:F1}
\end{center}
\end{figure}

\begin{figure}[tbp]
\begin{center}
\includegraphics[height=5cm,width=6cm]{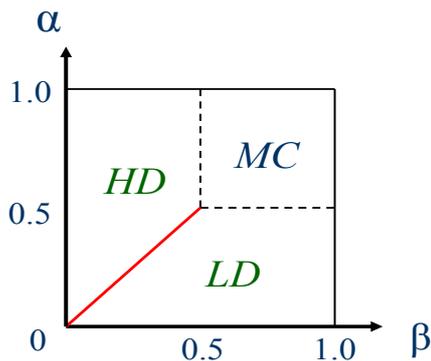}
%\hspace{-1cm}
\caption{Phase diagram of the proto TASEP. Maximal current and
high/low density phases are denoted by MC, HD, and LD respectively. The
transition between the MC phase and the other two (dashed lines) is
continuous. The transition across the HD-LD boundary (solid, red online) is
discontinuous.  }
\label{fig:F2}
\end{center}
\end{figure}

In this model, the sites of a one-dimensional lattice, labeled by $i=1,...,L$%
, may be occupied by $n_i=0$ or $1$ particle, so that a configuration of the
system is specified by $\left\{ n_i\right\} $. Starting from some initial
configuration with $N<L$ particles, the rules of evolution are the
following. At each update step, a site is randomly chosen and, provided it
is occupied and the next site, located in the selected direction, is empty,
an attempt to exchange the particle-hole pair is made. With probability $%
\gamma $, the attempt is termed ``successful'' and the particle is moved to
the next site. For the process on a ring, site $L$ is connected to site $1$.
Then, a Monte Carlo step (MCS) is defined as $L$ attempts, so that, on the
average, each site will be chosen once in a MCS. This way, it makes sense to
compare systems with very different $L$'s, but run for the same number of
MCS. For the open TASEP, two additional rules apply: Site $1$, if empty, is
filled with probability $\alpha $ and a particle on site $L$ leaves the
lattice with probability $\beta $. An MCS in this case consists of $L+1$
attempts. For the rest of this article, we will focus on the open case,
since it resembles the process of protein synthesis. Schematically, its
rules are summarized in Fig.~\ref{fig:F1}.
%================================
%================================================
The central question is: What is $P\left( \left\{ n_i\right\} ,t\right) $,
the probability for finding the system in configuration $\left\{ n_i\right\} 
$ after $t\,$ attempts? A master equation for $P$ can be easily written (for
open TASEP):
\begin{equation}
P\left( \left\{ n_i^{\prime }\right\} ,t+1\right) -P\left( \left\{
n_i^{\prime }\right\} ,t\right)  =\sum_{\left\{ n_i\right\} }{\cal L}%
\left( \left\{ n_i^{\prime }\right\} ,\left\{ n_i\right\} \right) P\left(
\left\{ n_i\right\} ,t\right)   \label{ME} \\
\end{equation}
where
\[
{\cal L}\left( \left\{ n_i^{\prime }\right\} ,\left\{ n_i\right\} \right)
\equiv W\left( \left\{ n_i^{\prime }\right\} ,\left\{ n_i\right\} \right)
-\delta \left( \left\{ n_i^{\prime }\right\} ,\left\{ n_i\right\} \right)
\sum_{\left\{ n_i^{\prime \prime }\right\} }W\left( \left\{ n_i\right\}
,\left\{ n_i^{\prime \prime }\right\} \right) 
\]
is known as the Liouvillian (which plays a role similar to the Hamiltonian
in quantum mechanics) and $\delta $ is the Kronecker delta. Here, $W\left(
\left\{ n_i^{\prime }\right\} ,\left\{ n_i\right\} \right) $ is the
transition probability from $\left\{ n_i\right\} $ to $\left\{ n_i^{\prime
}\right\} $%
\begin{eqnarray}
&&\frac 1{L+1}\left[ \alpha \left( 1-n_1\right) \delta \left( n_1^{\prime
},n_1+1\right) \prod_{j>1}\delta \left( n_j^{\prime },n_j\right) \right.  
\nonumber  \label{w1} \\
&&+\sum_{k=1}^{L-1}\gamma n_k\left( 1-n_{k+1}\right) \delta \left(
n_k^{\prime },n_k-1\right) \delta \left( n_{k+1}^{\prime },n_{k+1}+1\right)
\prod_{j\neq k,k+1}\delta \left( n_j^{\prime },n_j\right)   \nonumber \\
&&\left. +\,\,\beta n_L\delta \left( n_L^{\prime },n_L-1\right)
\prod_{j<L}\delta \left( n_j^{\prime },n_j\right) \right] \,  \label{W}
\end{eqnarray}
where the changes $n\rightarrow n\pm 1$ are explicitly displayed. However,
finding the solution to (\ref{ME}) is far more difficult. A simpler
question is: What is the stationary distribution, $P^{*}\left( \left\{
n_i\right\} \right) $, assuming the system settles into such a $t$%
-independent state at large times? Once it is known, other natural questions
arise: What are the macroscopic properties of the system in this state? Of
particular interest is how $\alpha $ and $\beta $ control the averages of
observables, 
\[
\left\langle {\cal O}\right\rangle \equiv \sum_{\left\{ n_i\right\} }{\cal O}%
\left( n_i\right) P^{*}\left( \left\{ n_i\right\} \right) \,\,, 
\]

such as the density profile $\rho _i\equiv \left\langle n_i\right\rangle $.
Once we have $\rho _i$, other quantities of interest can be computed. In
particular, we will be mostly interested in the overall density, $\bar{\rho}%
\equiv \sum_i\rho _i/L$, and the average current, $J=\beta \rho _L$ (i.e.,
the average number of particles that enters/exits the lattice in a Monte
Carlo step). The answers, some known to Gibbs, {\em et. al.}, became more
well-established over the last two decades. Setting $\gamma $ to unity
(without loss of generality), the system can be found in three distinct
phases in the $\alpha $-$\beta $ plane -- a half-filled phase with maximal
current and high/low density phases \cite{Krug91}, denoted by MC, HD, and LD
respectively in Fig.~\ref{fig:F2}. 
%================================
%================================================

Transitions from the maximal current (MC) phase to the other two phases 
(HD/LD) are continuous, and display critical behavior similar to
second order phase transitions in equilibrium. Indeed, critical properties,
such as algebraic decaying correlations, can be found in the entire MC
phase. Across the HD-LD boundary, the transition is discontinuous, and, on
the line itself, the system displays coexistence of HD and LD. Specifically,
the HD-LD regions are macroscopic, separated by a microscopic interface,
referred to as a ``shock.'' As in many equilibrium systems with coexistence,
such an interface can be located anywhere. In TASEP, the shock performs a
random walk (reflected only from the ends), so that the average density
profile is linear in $i$, interpolating between $\bar{\rho}_{HD}$ and $\bar{%
\rho}_{LD}$. In the literature, this line is often referred to as the ``shock
phase'' (SP). Setting up a phenomenological theory for the behavior of this
shock, known as domain wall theory, several authors have been successful in
predicting many properties of TASEP {\em outside} the MC region 
\cite{BS02,SA02}. The exact $P^{*}\left( \left\{ n_i\right\} \right) $ was
found \cite{DDM92,DEHP93,SD93,Schutz93}, from which $J\left( \alpha ,\beta
\right) $ and $\bar{\rho}\left( \alpha ,\beta \right) $ can be computed
analytically for all $\alpha ,\beta $. In the $L\rightarrow \infty $ limit,
these are remarkably simple: $J=\bar{\rho}\left( 1-\bar{\rho}\right) $
always, while $\bar{\rho}=\left\{ 1/2,1-\beta ,\alpha \right\} $ in the
MC,HD,LD phases, respectively. Thus, $J\leq 0.25$ in general. More recently,
considerable progress was made using the powerful Bethe-Ansatz \cite
{KSKS98,DS00,NAS02,TMH03,GMGB07,dGE06,PPOF05}, so that the complete spectrum and all the
eigenvectors of ${\cal L}$ are accessible. Consequently, some of the more
complex, dynamic properties of TASEP are also exactly known. Details of this
large body of results are beyond the scope of this article. The interested
reader may consult several comprehensive reviews such as \cite{GM06,BE07,Derrida07}.

Despite this comprehensive knowledge of ${\cal L}$, there are seemingly
simple questions about this system for which simple answers are not
available. An example is the power spectrum associated with $N\equiv
\sum_{i}n_{i}$, the total number of particles on the lattice. Specifically,
we record a time series $N\left( t\right) $ over a run and construct its
Fourier transform, $\tilde{N}\left( \omega \right) $. Carrying out many runs
and taking the average, $\overline{\left| ...\right| }$, the power spectrum
is 
\[
I\left( \omega \right) \equiv \overline{\left| \tilde{N}\left( \omega
\right) \right| ^{2}}\,\,=\overline{\left| \frac{1}{T}\sum_{t=0}^{T}e^{i%
\omega t}N\left( t\right) \right| ^{2}}. 
\]
Note that this average contains information on the dynamics and is therefore
not related to the static average, $\left\langle ...\right\rangle $, above.
If the runs are taken when the system is in the steady state, then $I\left(
0\right) $ is, of course, known: $\left( \bar{\rho}L\right) ^{2}$. But, $%
I\left( \omega >0\right) $ displays more interesting behavior, such as
oscillations (in $\omega $) in the HD/LD phases \cite{AZS07,CZ10}. Although
the physics behind these is understandable and approximate theories provide
reasonable fits, an exact analytic formula is not known (except formally)\footnote{%
In this case, the difficulties lie mainly in computing the average of
nonlocal (in both space and time) operators.}. In the remainder of this
article, we will look beyond this proto model and focus on generalizations
which take into account some other essential ingredients in the process of
protein synthesis.

Before continuing, let us point out an equivalent formulation of the open
TASEP, but based on a ring. Conceptually simpler and essentially used in
simulations, this version will appear to be most natural in the contexts to
be presented below. Here, we consider a periodic lattice with $L+1$ sites
filled with a total number, $N_{tot}$, of particles. Considering the role it
plays, we will refer to the extra site, $i=0$, as the ``reservoir'' or the
``pool.''\footnote{%
The notion of particle reservoirs was used in the literature, with one major
difference. Unlike here, open TASEPs were coupled to {\em two} unrelated
reservoirs, one at each end.}The rules associated with this site are, of
course, quite different from those in the bulk: (i) It has unlimited
occupation, so that we are guaranteed $n_0\geq 1$ by imposing $N_{tot}\geq
L+1$. (ii) If it is chosen for updating, one of its particles is moved to $%
i=1$ with probability $\alpha \left( 1-n_1\right) $. (iii) If site $L$ is
chosen and $n_L=1$, the particle hops into the pool with probability $\beta $%
, regardless of $n_0$. By denoting $\alpha ,\beta $ as $\gamma _0,\gamma _L$%
, we may regard them as part of a full set of site-dependent hopping rates $%
\left\{ \gamma _i\right\} $. Incorporating the special rules for site $0$,
we can replace $\left[ ...\right] $ in (\ref{W}) by a succinct expression 
\begin{equation}
\sum_{k=0}^L\gamma _k\left[ n_k+\delta _{k,0}\left( 1-n_0\right) \right]
\left[ 1-n_{k+1}+\delta _{k,L}n_L\right] \prod_{j=0}^L\delta \left(
n_j^{\prime },n_j-\delta _{j,k}+\delta _{j,k+1}\right) \,\,.  \label{Wall}
\end{equation}
(with $n_{k+1}=n_0$, etc.) Note that this $\Pi _j\delta $ includes all the
possible changes in $\left\{ n_j\right\} $. To re-emphasize, $N_{tot}=n_0+N$
is conserved in this formulation. However, as long as $N_{tot}\geq L+1$, the
properties of the open TASEP above are identical to the $i\in \left[
1,L\right] $ part of our ring and $N$ can fluctuate in the range $\left[
0,L\right] $ as before.

\subsection{\label{general}Generalizations of TASEP}

\begin{figure}[tbp]
\begin{center}
\includegraphics[height=4cm,width=10cm]{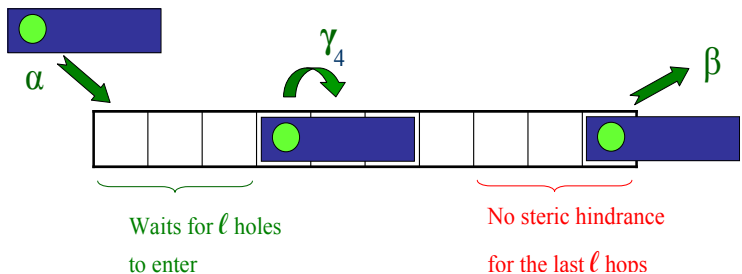}
\hspace{12cm}
\caption{ ``Complete
entry, incremental exit'' rule for an $\ell =3$ case. The gray dots (green online) denote
the ``readers.'' Since the reader of the second particle is on $i=4$, it
will hop with probability $\gamma _4$}
\label{fig:F3}
\end{center}
\end{figure}

As noted above, Gibbs, et. al. \cite{MG69,MGP68} were aware of the size of a
ribosome compared to a codon, so that Spitzer's simple TASEP must be
generalized to having ``particles'' which extend over $\ell \geq 1$ sites.
Indeed, from the latest data, $\ell \sim 12$ seems to be the most
appropriate \cite{HR80,MGP68,MG69} . This generalization requires
some modifications to the rules. Since the ribosome appears to ``read'' the
codon over its A-site, it is most natural to associate one particular part
of the extended particle with the ``reader'' \cite{SZL03}. After some
thought, it is also clear that, as far as TASEP is concerned, which part is
labeled the reader is irrelevant. For convenience, we choose the reader to
be at the trailing edge of the particle \cite{SZL03}. To ``read'' the first
codon (site $1$ on the lattice), the ribosome/particle must enter the
lattice and for that to occur, the first $\ell $ sites must be empty. On the
other hand, while the ribosome is ``reading'' the last $\ell $ codons, it
must be the last particle on the lattice, with no others to impede its
progress. Therefore, it can move without hindrance toward the exit end. The
new set of entry/exit rules is known as ``complete entry, incremental exit'' 
\cite{LC03}.(See Fig.~\ref{fig:F3} for an illustration.)
%================================
%================================================

These seemingly modest changes of
the rules have profound consequences. At the simplest level, we must now
distinguish between particle (ribosome) density and ``coverage'' density
(number of sites ``covered'' by a particle per unit length). Denoting the
former by $\rho ^r$ and the latter by $\rho $, we see that the overall
densities differ by a factor of $\ell $ 
($\bar{\rho}=\ell \,\overline{\rho ^r}$), and the two profiles are related by 
\[
\rho _i=\sum_{k=0}^{\ell -1}\rho _{i-k}^r\,\,. 
\]
If $\rho ^h=1-\rho $ denotes the hole density, we also have $\rho ^r+\rho
^h<1$. Although the phase diagram and the current-density relation $J\left( 
\bar{\rho};\ell \right) $ are qualitatively unchanged  
\cite{SZL03,LC03,DSZ07_PRE} , an exact solution 
(for $P^{*}$ or $J$ or $\bar{\rho}$) remains elusive.
Stationary profiles are much more seriously affected, especially in the HD
phase \cite{DSZ07_PRE}. On the other hand, for a\ homogeneous TASEP on a ring, $%
P^{*}$ is known to be uniform \cite{Spitzer70}, so that an exact $J\left( 
\bar{\rho};\ell \right) $ can be derived \cite{SZL03,LC03}\footnote{
Other exact results (also for a collection of particles with different $\ell 
$'s) have been found recently. See e.g., \cite{AL03}.}. 
To be precise, we denote the particle
current by $J$, but write it in terms of $\bar{\rho}$ instead of $\overline{%
\rho ^r}$: 
\begin{equation}
J\left( \bar{\rho};\ell \right) =\frac{\bar{\rho}\left( 1-\bar{\rho}\right) 
}{\ell +\left( \ell -1\right) \bar{\rho}}\,\,.  \label{J-rho-ell}
\end{equation}
A more elegant version of this formula is 
\begin{equation}
J\,^{-1}=\left[ \,\,\overline{\rho ^r}\,\,\right] ^{-1}+
\left[ \,\,\overline{\rho ^h}\,\,\right] ^{-1}\,\,,
\end{equation}
in which the second term accounts for steric hindrance somehow. 
 On the left
is the average time between successive particles (with exclusion) exiting
the lattice. On the right, we have the sum of such times for {\em %
non-interacting} particles and holes. This connection is quite remarkable.

Returning to eqn. (\ref{J-rho-ell}), we see that $J\left( \bar{\rho}\right) $
still rises from zero, reaches a maximum, and returns to zero. However, its
upper bound is lowered to $\left( 1+\sqrt{\ell }\right) ^{-2}$ , i.e., by $%
O\left( \ell \right) $ for large $\ell $ \cite{SZL03,LC03,DSZ07_PRE}. While this 
$J\left( \bar{\rho};\ell \right) $ it is not rigorously the same as the one
in an open TASEP, it can be argued that, in the $L\rightarrow \infty $
limit, the two should be the same. As noted above, Gibbs, et. al. arrived at
the same $J\left( \bar{\rho};\ell \right) $ long ago, by accounting for some
effects of the $\ell $-exclusion approximately. This is one of the few
reasonably well understood aspects of TASEP with extended objects. In
passing, we should mention that TASEPs with polydispersed particles on a
ring have also been studied \cite{AL03}, though their relevance to protein
synthesis seems remote.

A second essential aspect of our problem was also recognized by Gibbs, et.
al. \cite{MG69,MGP68}, namely, site-dependent hopping rates, i.e.,  
inhomogeneous TASEPs. In Section \ref{Rudi}, we indicated the rationale for considering
such a difficult problem: non-uniform aa-tRNA abundance. Needless to say, it
is prohibitively difficult to determine quantities like $J$ and $\bar{\rho}$
for a TASEP with an arbitrary set of rates, $\left\{ \gamma _i\right\} $.
Even when restricted to point particles ($\ell =1$) on a homogeneous ring,
the introduction of a single ``defective site'' (with $\gamma \neq 1$ but no
changes to the rules of exclusion) renders the problem insolvable (i.e., no
exact $P^{*}$) so far. The non-trivial consequences and serious challenges
were noted as early as 1992 \cite{JL92}.\footnote{%
Remarkably, exact results are available if a single {\em particle} hops more
``slowly'' \cite{JL92,DJLS93,Schutz93,Mallick96}.} With several
defects, systematic studies become less manageable, even with approximate or
numerical methods. For an open TASEP with a few defect sites, progress was
made mainly with Monte Carlo simulations, while some understanding is
possible by exploiting mean-field approximations of various levels \cite
{Kolomeisky98,FKT08,DZS09}. 
Most relevant to modeling translation is the discovery that the
current (for $\alpha =\beta =1$) depends on the location of slow defects ($%
\gamma <1$) \cite{CL04,DSZ07_JSP,DSZ07_PRE,FKT08,DZS09}. 
In particular, if there are two slow sites in the bulk, the
distance between them affects $J$ seriously 
\cite{CL04,DSZ07_JSP,DSZ07_PRE,FKT08}. This implies that protein production
rates can be significantly suppressed if codons associated with rare
aa-tRNA's are {\em clustered} in the gene. At the other extreme, several
groups studied TASEPs with a full set of quenched random rates, $\left\{
\gamma _i\right\} $, each of which is chosen from from some distribution
(e.g., Gaussian, two-valued, etc.) \cite{TB98,BBEM99,Krug00,HS04,GS08}.
These authors considered only point particles and focused on the effect of
disorder on the (quenched average) current-density relation. Using
simulations and mean field approximations, $J\left( \bar{\rho}\right) $ is
found to develop a plateau in a region around $\bar{\rho}=1/2$, details of
which depend on the variance of the distribution of the inverse rates: $%
1/\gamma $. The phase diagram remains qualitatively the same, with three
phases that resemble MC, HD, and LD. Not surprisingly, the main effect is
that the transitions are no longer sharp. Beyond these studies, disordered
TASEPs with $\ell >1$ are yet to be explored.

To model protein synthesis more realistically, we need a combination of at
least three ingredients: (a) open boundary conditions, (b) extended objects
(say, $\ell =12$), and (c) inhomogeneous rates, $\left\{ \gamma _{i}\right\} 
$. As we noted from the historic perspective, it took some time to arrive at
full solutions -- for TASEPs with ingredients (a) or (b). Yet, here we would
prefer to include all three aspects and ask for, at the least, the average
current $J\left( \alpha ,\beta ,\left\{ \gamma _{i}\right\} ,\ell \right) $
and the overall density $\bar{\rho}\left( \alpha ,\beta ,\left\{ \gamma
_{i}\right\} ,\ell \right) $. Clearly, this program is extremely ambitious,
even if we restrict our investigations to the Monte Carlo approach. In
Section \ref{simple}, we will present a very simple, yet reasonably reliable, method
to arrive at a good estimate for the current.

\section{\label{recent}Some recent developments}

In this section, we present two topics where some recent progress was made.
We begin (Section \ref{compete}) with an analysis of a particular instance of the cell
having limited resources available. In TASEP language, we are exploring how
a TASEP is affected by having a {\em finite} reservoir of particles. The
effects of several TASEPs competing for the same pool of particles \cite
{ASZ08,CZ09,CZS09} will be also presented. The rationale behind such
pursuits is that a cell has thousands of {\em copies} of thousands of
different {\em types} of mRNAs, competing for the same pool of ribosomes.
Do some ``win'' while others ``lose''? In the language of TASEPs, since we
model an mRNA by a sequence $\left\{ \gamma _i\right\} $, we will be
interested in $J\left( \left\{ \gamma _i\right\} ;N_{tot}\right) $, namely,
how the current associated with this sequence depends on $N_{tot}$, the
total number of particles in the pool. Our analysis here will be restricted
to homogeneous TASEPs ($ \gamma _i = 1 $) of differing lengths.

Section \ref{simple} IVb will be devoted to our search for a simple (``quick and dirty'')
way to estimate $J\left( \alpha ,\beta ,\left\{ \gamma _{i}\right\} ,\ell
\right) $ for a single TASEP but with a fully inhomogeneous sequence of
hopping rates, $\left\{ \gamma _{i}\right\} .$ This search leads us to a
novel form of quenched randomness, which we named\footnote{This notion was presented in \cite{APS09}.} 
``distribution of distributions.'' Recall that a protein is a fixed
sequence of $L$ amino acids and can be coded by $O\left( e^{L}\right) $
different mRNAs. Suppose we wish to synthesize an artificial protein
consisting of only R's. We can use any one of the $6^{L}$ possible mRNAs,
each of which corresponds to a realization of a quenched random sequence of
codons, chosen from a single distribution of 6 values (corresponding to CGU,
CGC, CGA, CGG, AGA, or AGG in this example). This procedure is standard for
problems involving quenched disorder\footnote{See, e.g., \cite{EA75,YY98,Dotsenko01}, etc.}. 
However, in a
naturally occurring protein -- the ``wild type'' -- the $L$ amino acids will
be different, so that the sequence of degeneracies will be non-trivial and
fixed (e.g., 4266224 for the amino acid string PQLRFEV). Thus, instead of
choosing codons from a single distribution to construct all possible mRNAs
(as in the artificial RRRRR case or in all previous studies of quenched
disorder), they must be chosen from a fixed sequence of {\em different}
distributions. Pursuing these ideas further, we discovered a remarkable fact
about {\em E. coli}. Simulating 5000 randomly chosen sequences for each of
10 specific genes, we find that the average currents lie in a narrow range
(within 25\% of each other). However, the currents associated with the wild
types typically lie very far above the average. These are intriguing
findings, from the perspectives of both, the statistical physics of quenched
disorder and the specific realization ``chosen''\ by the living organism.

\subsection{\label{compete}Competition for ribosomes: TASEP with finite particle
reservoirs}

In a living cell, ribosomes are constantly synthesized and degraded. On the
other hand, it is believed that some are also ``recycled,'' i.e., after
termination in translating one gene, the subunits reassemble to translate
another gene. Of course, there are multiple facets to ``ribosome
recycling.'' Chou considered the enhancement of initiation rates on a gene
due to the proximity of a ribosome which unbinds from the same mRNA \cite
{Chou03}. We consider a different aspect. Ignoring synthesis/degradation, let
us model the number of ribosomes in a cell by a constant, $N_{tot}$, to be
shared by all the genes. Then we ask: What is the effect of multiple TASEPs
competing for a single pool with a finite number of particles? As a
base-line study, we first focus on the effects of finite $N_{tot}$'s on just
one homogeneous TASEP \cite{ASZ08,CZ09,CZ10}. For example, we seek $\bar{\rho%
}\left( N_{tot}\right) $, the dependence of the overall density on the total
particle number. This study is then extended to include multiple
(homogeneous) TASEPs \cite{CZS09} with possibly different $L$'s. Will the
overall densities and currents be the same or different? If the latter, how
are they controlled by $N_{tot}$? So far, all studies are based on point
particles and uniform entry/exit rates.

For these investigations, it is clear that the alternative representation of
an open TASEP in Section \ref{proto} is most natural, with $n_{0}$ being the number
of particles in the pool. In the single TASEP case, novel behavior already
arises when we introduce only one modification: allowing $N_{tot}$ to be
lowered below $L+1$. Ha and den Nijs coined this the ``parking garage
problem'' and provided many interesting results \cite{HdN02}. To model how
translation might be affected by the scarcity of ribosomes, we let the
binding rate of a ribosome to the mRNA, $\gamma _{0}$, depend on the
ribosome concentration. In particular, when the ribosome concentration is
very low, we let $\gamma _{0}$ be proportional to it. At the opposite
extreme, it should have no effect on $\gamma _{0}$, which should take on
some intrinsic value -- denoted by $\alpha $ -- associated purely with the
binding kinetics. In our model, $n_{0}$ is proportional to this
concentration, so that we simply choose a convenient $\gamma _{0}\left(
n_{0}\right) $ which interpolates between $0$ and $\alpha $. In all the
simulation studies \cite{ASZ08,CZ09,CZ10}, we have 
\begin{equation}
\gamma _{0}\left( n_{0}\right) =\alpha \tanh \left( n_{0}/N^{\ast }\right)
\end{equation}
where $N^{\ast }$ is some crossover parameter (chosen to be $O\left(
L\right) $ for convenience). By contrast, the exit rate should not be
affected by $n_{0}$, so that we simply have $\gamma _{L}=\beta $. Since $%
N_{tot}=n_{0}+N$, both $n_{0}$ and $N$ will be small as $N_{tot}$ is
increased from $0$, and the system first finds itself in the LD phase. At
the other extreme, $n_{0}$ is necessarily large as well (since $N\leq L$),
so that we will arrive at an ordinary open TASEP associated with ($\alpha
,\beta $). A crossover occurs when $N_{tot}$ reaches $O\left( N^{\ast
}\right) =O\left( L\right) $. As may be expected, the LD-LD and LD-MC
crossovers are uneventful, since no discontinuities are encountered. The
response of the TASEP can be well approximated by a self-consistent equation
for $\bar{\rho}\left( N_{tot}\right) :$%
\begin{equation}
\bar{\rho}=\gamma _{0}=\alpha \tanh \left( \left( N_{tot}-\bar{\rho}L\right)
/N^{\ast }\right)  \label{LD-rho}
\end{equation}

More interesting is the LD-HD crossover, since it spans a discontinuous
boundary. The response is well described by the following. Raising $N_{tot}$
from $0$, the average density is given by the above equation, until a
critical value, $N_{tot}^{-}\equiv \beta L+N^{\ast }\tanh ^{-1}\left( \beta
/\alpha \right) $, is reached. Lowering $N_{tot}$ from $\infty $, $\bar{\rho}
$ remains at the HD value of $\bar{\rho}_{HD}\equiv 1-\beta $, until $%
N_{tot} $ reaches another critical value: $N_{tot}^{+}\equiv $ $\left(
1-\beta \right) L+N^{\ast }\tanh ^{-1}\left( \beta /\alpha \right) $.
Between $N_{tot}^{-}$ and $N_{tot}^{+}$, all increases in $N_{tot}$ are
absorbed by the TASEP (while $n_{0}$ and $\gamma _{0}$ stay constant). Thus, 
$\bar{\rho}$ rises {\em linearly:} $\bar{\rho}\left( N_{tot}\right) =\beta
+\left( N_{tot}-N_{tot}^{-}\right) L$. Such a response has an analog in
equilibrium first order transitions, corresponding to, e.g., the linear
section in an isotherm in the $P$-$V$ diagram of a binary mixture.
Furthermore, the average profile ($\rho _{i}$) in this regime is also
noteworthy. Instead of being linear in $i$ (as in an unconstrained TASEP),
it resembles a stationary shock. The underlying physics is understandable:
The feedback from the pool prevents the shock from wandering throughout the
lattice. Instead the shock is localized to a position controlled by $N_{tot}$
while its fluctuations are controlled by another detail of the feedback: $%
\partial \gamma _{0}/\partial n_{0}$. Domain wall theory, so successful in
providing good approximations for an ordinary TASEP, can be generalized to
account for the feedback to give excellent ``zero-parameter fits'' to
simulation data \cite{CZ09}. In passing, let us mention that even more
remarkable structures appear in the special case of LD-SP (i.e., setting $%
\alpha =\beta $ and varying $N_{tot}$). In all cases, the current displays
no major surprises, mainly following the $J\left( \bar{\rho}\right) $ curve
of an unconstrained TASEP. The interested reader is referred to \cite{CZ09}
for details.

Next, we turn to multiple TASEPs and their competition for a finite pool of
particles \cite{CZS09}. To model different genes and the many possibilities
of regulation, we need (at the least) three parameters for each type 
($\mu $) of TASEP: $L_{\mu }$, $\alpha _{\mu }$, $\beta _{\mu }$. A 
systematic study in the full parameter space of $M$ TASEPs becomes quickly
unmanageable, so ours is restricted to $M=2,3$. On the other hand, as an
attempt at more realistic models, one unpublished study \cite{Cook10}
simulated the competition of 10 genes from {\em E.coli} (with $L$'s ranging
from 109 to 558; details in the next subsection). With $\ell =12$ and the
appropriate sets of $\gamma _{i}$'s, the only unrealistic part of this study
is setting $\alpha _{\mu }=1$ for all genes. Not surprisingly, for $%
N_{tot}\sim O\left( 1\right) $, the currents are all the same, being
controlled by the same small entry rate: $\gamma _{0}$. For $N_{tot}\gtrsim
300$, each TASEP is saturated in their MC-like phases, so that the currents
differ by a factor of \symbol{126}2. The approximate value of 300 can be
expected from 10(genes)$\times $400(typical length)$/$12($\ell $).
Meanwhile, crossovers occur at $N_{tot}$'s in the range of 100-250. The
conclusion of this limited study is that, while the first attempt has been
made at the question of mRNA competition for ribosome in a real cell, much
more remains to be explored before meaningful insights can be developed.

\begin{figure}[tbp]
\begin{center}
\includegraphics[height=10cm,width=7.5cm]{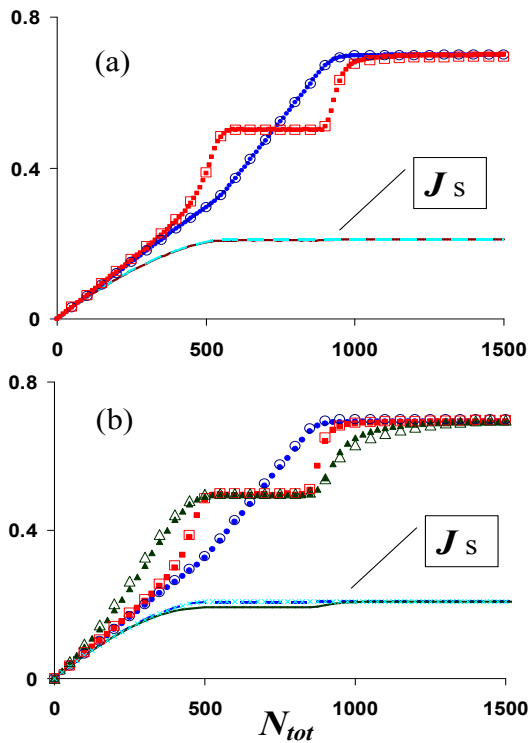}
\caption{ Average overall densities and currents as a function of $N_{tot}$ 
when two/three TASEP are competing for a single pool of particles. 
In both cases, $\alpha =0.7$ and $\beta =0.3$. The currents in all cases follow
approximately the same curve of the single TASEP case. Denoted by
essentially indistinguishable solid and dashed lines (color online), they are
marked by the call-out ``$J$s''. Simulations (solid symbols) and predictions from 
generalized domain wall theory (open symbols) of the densities are as follows.
(a) $L_1=1000$ (circles, blue online) and $L_2=100$ (squares, red online). 
(b) $L_1=1000$, $L_2=100$, and $L_3=10$ (triangles, green online). }
\label{fig:F4}
\end{center}
\end{figure}

Focusing on a more systematic (though less ``realistic'') study of
competition, we consider two TASEPs. The model here consists of two lattices
with $L_{1,2}$ sites, joined at one site (site $0$, the pool), so that it
has the topology of two rings joined at one point. When site $0$ is chosen,
with equal probability a particle attempts to move onto one of the two
lattices. Once a lattice is chosen, it enters with the rates, $\alpha _{1,2}$%
. As usual, there is no exclusion at site $0$, so that particles simply hop
from sites $L_{1,2}$ into the pool with rates $\beta _{1,2}$. For
simplicity, we let $\alpha _{1,2}=\alpha $ and $\beta _{1,2}=\beta $ in this
initial study \cite{CZS09}. Perhaps to be expected, when the two TASEPs are
identical (i.e., $L_1=L_2$), the symmetry is not spontaneously broken. The
two response curves are the same, within statistical fluctuations. However,
when the lengths are very different, a new pattern emerges. In particular,
for $L_1=1000$ and $L_2=100$, we find roughly {\em five} regimes in the LD-HD
crossover. While the longer TASEP displays essentially the same behavior as
in the single TASEP case (three regimes), the shorter one experiences more
variety (Fig.~\ref{fig:F4})

%================================
%================================================

It is remarkable that, in the central section, 
$\bar{\rho}\left( N_{tot}\right) $ for both are linear, with the shorter one
being a constant! It turns out that the shock in this TASEP is delocalized,
but acts as a control for the shock in the longer one. The motion of the two
shocks is completely {\em anti}-correlated, so that $n_0$, the pool particle
number, remains essentially fixed. As a result, the average density profiles
are quite different, being strictly linear for the short one. For the longer
TASEP, the profile can be readily described by the profile of a single
constrained TASEP, but ``smeared out'' over a distance of $L_2$ (length of
the shorter lattice). As can be seen from Fig.~4a, the generalized domain
wall theory is quite successful at capturing all this novel behavior.
Finally, these insights can be exploited to understand the behavior of three
TASEPs in competition. Though not dramatically different, new features do
appear, especially in cases where the lengths are widely separated. For
example, Fig.~4b is an illustration of $\bar{\rho}\left( N_{tot}\right) $
for $L_1=10$, $L_2=100$, and $L_3=1000$. Many other results, such as remarkable
properties of the stationary $P^{*}\left( \left\{ k_\mu \right\} \right) $,
the probability to find the domain wall located at site $k_\mu $ in lattice $%
L_\mu $, are available. Beyond the scope of this article, these may be found
in \cite{CZS09}. Of course, we have taken only a minuscule step towards
modeling competition in real cells. In addition of containing thousands of
different genes (e.g., $5416$ in one strain of {\em E. coli}), there can be
thousands of copies of each type, not to mention that we should include the
three essential ingredients pointed out at the end of Section \ref{TASEP}. Finally,
looking far ahead, we can consider the genes competing for finite pools of
the 46 varieties of aa-tRNA, a problem involving feedback from the details
of the average occupation at each site. Clearly, this is a gargantuan task
and much remains to be investigated before we can claim to understand
competition for finite resources in a real cell.

\subsection{\label{simple}A simple estimate for currents in the inhomogeneous TASEP}

In this subsection, we return to a single open TASEP with extended particles
($\ell >1$) hopping along with a fully inhomogeneous set of rates $\left\{
\gamma _i\right\} $. We will focus only on the average current $J\left(
\alpha ,\beta ,\left\{ \gamma _i\right\} ;\ell \right) $. The task of
predicting this $J$ is clearly beyond our present analytic abilities. Faced
with this impasse, one reasonable question is: Is there a simple way to
estimate it? Of course, in the limit of $\alpha \ll \gamma _{i>0}$, the
particle density will be exceedingly low so that the particles are
non-interacting, to a good approximation. Then, we simply have $J\simeq
\alpha $, since the total time it takes a particle to traverse the lattice
(mRNA) becomes irrelevant\footnote{%
Actually, many exact results exist for the full stochastic problem of just a
single particle hopping on such a ring. But we will not pursue this line
further, since our main interest will be the many-body problem on open
lattices.}. Note that the exclusion plays no role, so that $\ell $ is also
irrelevant. This consideration, based on the idea of the ``worst
bottleneck,'' can be used to provide the most naive estimate of $J$: 
\begin{equation}
J_{worst\,\,b^{\prime }neck}\left( \alpha ,\beta ,\left\{ \gamma _i\right\}
\right) \thicksim \gamma _{\min }  \label{J-gmin}
\end{equation}
where $\gamma _{\min }$ is the minimum in the set $\left\{ \gamma _i\right\} 
$. Another possible estimate is to use the averages and variances of the
entire set $\left\{ \gamma _i^{-1}\right\} $, which we denote by $\bar{\eta}$
and $\sigma _\eta ^2$, respectively (following the notation in \cite{HS04}).
These quantities proved quite successful in the analysis of quenched random
averages of currents \cite{TB98,Krug00,HS04}. However, there are limitations
for both estimates when addressing issues of interest here, namely, finding
a reasonably good estimate of the current for a {\em specific }sequence $%
\left\{ \gamma _i\right\} $. As in the estimate $J\simeq \alpha $, (\ref
{J-gmin}) is useful only when the bottleneck is very severe. But, realistic
rates are typically not so extreme that $\gamma _{\min }$ is drastically
smaller than the rest of the $\gamma $'s. More crucially, if we have more
than one site with $\gamma _{\min }$, then $J$ will be affected by their
locations. For example, studies of just two slow sites 
\cite{CL04,DSZ07_PRE,FKT08} showed
that, having them as neighbors as opposed to being far apart, $J$ can be
lower by as much as a factor of $2$. Indeed, if the sequence contains a
consecutive string of $k$ such sites with $k\gg \ell $, then we can regard
this stretch as an open, homogeneous TASEP in its MC phase. The
considerations around eqn. (\ref{J-rho-ell}) then provide us with 
$J\thicksim $ $\gamma _{\min }\left( 1+\sqrt{\ell }\right) ^{-2}$. On the
other hand, if these $k$ slow sites are very far apart, then the estimate
for $J$ due to a {\em single} slow site 
\cite{CL04,SKL04,DSZ07_JSP,DSZ07_PRE,Kolomeisky98} 
(which reduces to $J_{worst\,\,b^{\prime }neck}$, to lowest order in 
$\gamma_{\min }$) should suffice. 
Thus, the clustering of many slow sites indeed suppresses $J$,
by as much as a factor of $20$ for $\ell =12$. Similar limitations for the
other estimate exist. Given a particular set $\left\{ \gamma _i\right\} $
(e.g., a real gene found in nature), we may compute $\bar{\eta}$ and $\sigma
_\eta ^2$ by {\em assuming} that this set is a good representative of the
underlying distribution of $\gamma $'s. Yet, neither of these quantities
contains any information on the {\em location} of slow sites. Thus, we face
quite a range of uncertainties when attempting to provide a good estimate.
In the remainder of this article, we propose a rough and simple, yet
tolerably reliable, estimate for 
$J\left( \alpha ,\beta ,\left\{ \gamma_i\right\} \right) $.

Since clustering of slow sites appears to play an important role, our
attempt is to consider a ``coarse-grained'' set of rates. In particular, we
follow the notion introduced in \cite{SKL04} and define 
\begin{equation}
\left( K_s\right) _i\equiv \left[ \frac 1s\sum_{k=0}^{s-1}\frac 1{\gamma
_{i-k}}\right] ^{-1}\,\,.  \label{Km}
\end{equation}
The sum in this expression is recognizable as the typical time for a (free)
particle to traverse a stretch of $s$ sites before site $i$. Thus, $\left(
K_s\right) _i$ can be regarded as a ``coarse-grained'' rate associated with
hopping from site $i$. Obviously, by setting $s=i=L$ we recover a quantity
that resembles $\bar{\eta}$, but our interest is more mesoscopic, e.g., $%
s\thicksim \ell $, since that would account for some of the effects of
clustering of slow sites. Combining this notion with the idea of the
bottleneck being the limiting factor, we propose that $K_{\ell ,\min }$, the
smallest rate in $\left\{ \left( K_\ell \right) _i\right\} $, can be
exploited to give a good estimate for $J$. Note that we are {\em not}
proposing $J\cong K_{\ell ,\min }$, since the (maximal) current for a
homogeneous TASEP would be $K_{\ell ,\min }\left( 1+\sqrt{\ell }\right) ^{-2}
$, a value 20 times lower than $K_{\ell ,\min }$ in the case of $\ell =12$!
Instead, our hope is that a linear relationship $J\propto K_{\ell ,\min }$
would be adequate. To be specific, let us focus only on $\ell =12$ TASEPs
with large entry/exit rates, i.e., $\alpha =\beta =\gamma _{\max }$, and use
Monte Carlo techniques to find $J\left( \gamma _{\max },\gamma _{\max
},\left\{ \gamma _i\right\} ;12\right) $. To simplify notation, this average
current will be denoted simply by $J\left( \left\{ \gamma _i\right\} \right) 
$. Then, allowing a phenomenological slope, $A$, we will test how well 
\begin{equation}
J\left( \left\{ \gamma _i\right\} \right) \simeq AK_{12,\min }
\label{J-Kmin}
\end{equation}
is obeyed.

\begin{figure}[tbp]
\begin{center}
\includegraphics[height=3.5cm,width=11.5cm]{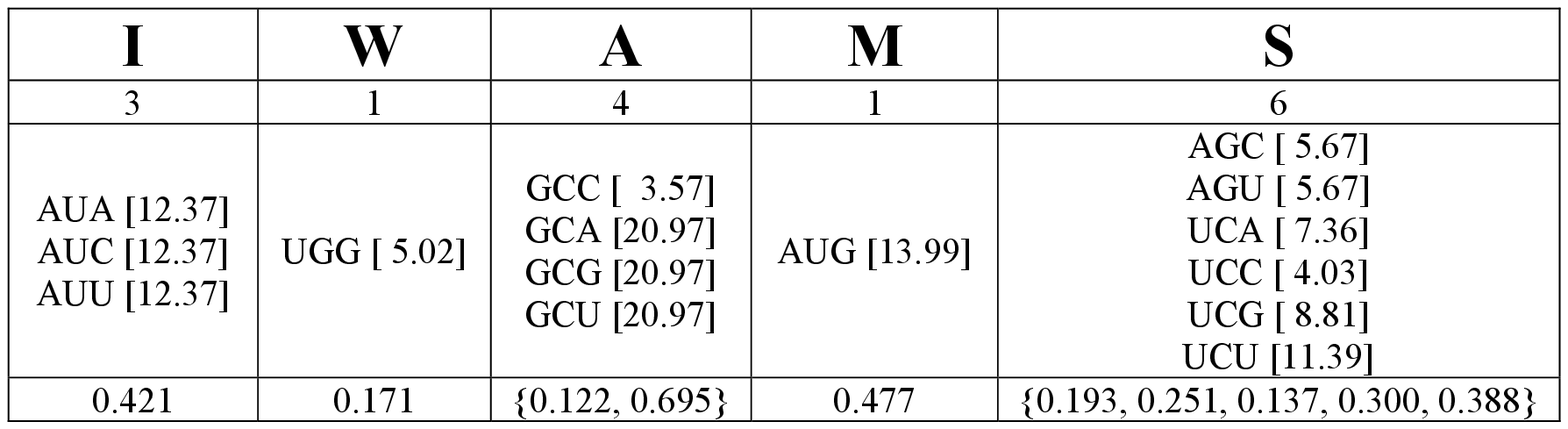}
\caption{A 5-amino acid ``designer gene'' IWAMS with its 
associated degeneracies ($m$) in the second row. The third row shows 
explicitly  the $m$ 
synonymous codons and aa-tRNA cellular concentrations from \cite{DNK96}.
The last row are the corresponding hopping rates used in our simulations, 
defined in eqn. (\ref{gammas}). }
\label{fig:F5}
\end{center}
\end{figure}

Before describing the results of such a test, let us provide some details on
the ensemble of genes we will use, as well as the concept of a quenched
``distribution of distributions.'' As pointed out above, if we wish to
synthesize a particular protein (a specific sequence of $L$ amino acids: $%
\left\{ aa_{i}\right\} $), we can use the codes from 
$\prod_{i}m\left( aa_{i}\right) $ 
different mRNAs, using the appropriate degeneracies given
in (\ref{aa-deg}). To help the readers, let us provide a simple example: a
fictitious $L=5$ ``protein chain,'' IWAMS, shown in the first row of 
Fig.~\ref{fig:F5}. From (\ref{aa-deg}), we see that the sequence
of $m$'s is 31416, shown in the second row. 
So, there are 72 (=3$\cdot $1$\cdot $4$\cdot $1$\cdot $6)
possible ``genes'' which can code for this ``protein.''  
All the possible codons are shown in the third row, so all 72 can
be read off, e.g., AUCUGGGCCAUGUCC. 

%================================
%================================================

One natural question is: If these 72
possibilities are generated with equal probability, what is the distribution
of the currents? To answer this, we must deal with another complication.
Corresponding to each codon is an aa-tRNA. But, their relative abundances
are not unique. Instead, many are the same (in {\em E.coli}, for a certain 
growth condition \cite{DNK96}), as shown within [...] next to each codon 
in the third row. In our
simulations, we normalized the hopping rate associated with the highest
abundance -- 29.35 -- to unity, and so, in the fourth row, we list all the
possible $\gamma $'s. So, of the 72 possible ``genes,'' there are only 10 (=1%
$\cdot $1$\cdot $2$\cdot $1$\cdot $5) distinct sequences of $\left\{ \gamma
_{i}\right\} $. Therefore, we should alert the reader to another
complication when considering our ensemble of 72 (equally probable)
``genes.'' Not all $\left\{ \gamma _{i}\right\} $'s are equally probable,
since there are only 10 possible {\em distinct} $\left\{ \gamma _{i}\right\} 
$'s. As an illustration, the set $\left\{
0.421,0.171,0.695,0.477,0.388\right\} $ is three times more likely to occur
as $\left\{ 0.421,0.171,0.122,0.477,0.388\right\} $, since 3 codons out of 4
coding for A have the same abundance, 20.97. Since average production rates
(i.e., $J$'s) depend only on the sequence $\left\{ \gamma _{i}\right\} $,
the probabilities of any $J$ occurring in our ensemble will not be uniform.

With this illustration in mind, let us define our notations for an explicit
formulation.

\begin{itemize}
\item  Let $\nu =1,2,...$ label the various proteins in a cell. Typically,
there would be thousands. Below, we will study just 10 in {\em E.coli}: five
highly expressed ones (dnaA, ompA, rspA, rplA, tufA) and five poorly
expressed ones (araC, lamB, lacI, secD, trpR). Each is a specific sequence
of $L_\nu $ amino acids, which we denote by $\left\{ aa_i\right\} _\nu
;\,\,i=1,...,L_\nu $. Of course, $1\leq aa\leq 20$, associated with the 20
alphabets in the first row of table (\ref{aa-deg}).

\item  To synthesize each $\nu $, there are $M_\nu $ distinct mRNAs
(sequences of codons), which we denote as $\left\{ c_i\right\} _\nu $. Let
us label these sequences by $\mu =1,...,$ $M_\nu $. Obviously, all sequences 
$\left\{ c_i\right\} _\nu $ has the same length as $\left\{ aa_i\right\}
_\nu $ so that $L_\mu =L_\nu $. Here, the variable $c$ lies between $1$ and $%
61$, but $c_i$ (its value at site $i$) is linked to the value of $aa_i$ (via
the aa-codon mapping). Recall that this mapping is one-to-many ($1$-$m_{aa}$%
), as the first three rows in table (\ref{aa-deg}) illustrate. Thus, $M_\nu
=\prod_{i=1}^{L_\nu }m_{aa_i}$ is large number, typically $O\left( \exp
L_\nu \right) $.

\item  Depending on the conditions in which a cell finds itself, different
aa-tRNAs are found with varying concentrations (e.g., ref. \cite{DNK96} for 
{\em E. coli}). Following typical notation, we write $\left[ c\right] $ for
the concentration of the aa-tRNA associated with codon $c$. As we see in the
third row of table (\ref{aa-deg}), the $c$-$\left[ c\right] $ mapping is
also often one-many. For simplicity, we assume the ribosome's hopping rate,
from site $i$ to $i+1$, to be proportional to $\left[ c_i\right] $.
Normalizing these rates so that unity is associated with the largest
concentration, $\left[ \max \right] $, we use 
\begin{equation}
\gamma _i\equiv \frac{\left[ c_i\right] }{\left[ \max \right] }
\label{gammas}
\end{equation}
for our simulations.
\end{itemize}

\begin{figure}[tbp]
\begin{center}
\includegraphics[height=9cm,width=9cm]{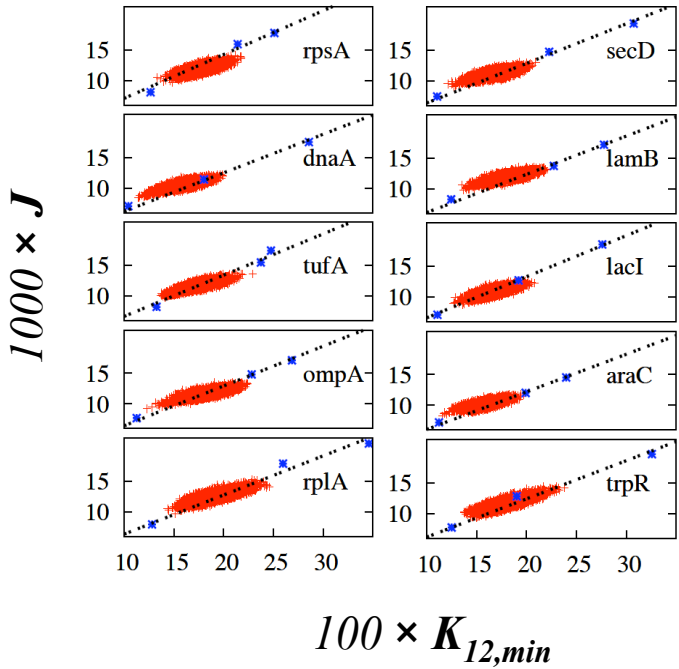}
\caption{Relation between $J$ and $K_{12, min}$ for synthesizing 10 
proteins in  {\em E.coli}. Stars ($\ast$, blue online) are from the 
``abysmal'', wild type and ``optimal'' sequences. The dash line 
(blue online) is the best linear fit through them and the origin. 
The elliptical cluster (red online) is from 5000 randomly 
generated sequences which code for the 10 indicated proteins. }
\label{fig:F6}
\end{center}
\end{figure}

\noindent With this framework in place, let us discuss our findings from
performing the following simulations. For each of the 10 proteins shown
above, we generated 5000 sequences $\left\{ c_{i}\right\} _{\nu }$ with no
bias, and compiled the associated $\left\{ \gamma _{i}\right\} _{\nu }$
accordingly. For each member in this ensemble, we computed $K_{12,\min }$
and simulated the associated TASEP (with $\ell =12$ particles) to obtain its
current $J$ $\left( \left\{ \gamma _{i}\right\} _{\nu }\right) $. These
pairs of values are plotted in the $J$-$K$ plane. They generally form an
elliptical cluster (5K indiscernible points, red online), as shown in 
Fig.~\ref{fig:F6} for each of the 10 proteins. 

%================================
%================================================

Two other features appear in these plots: a
dashed line and three points (stars), all being blue online. The lowest
point corresponds to an ``abysmal'' sequence, formed by having the lowest
allowed $\gamma $ at each site. Thus, it produces the lowest possible $%
\left( J,K\right) $. Similarly, the highest point is associated with an
``optimal'' sequence, with the highest possible $\left( J,K\right) $ for
this protein. The point in the middle is derived from the wild type
(naturally occurring) sequence. Finally, the dashed line is the best linear
fit through the three points, {\em constrained }to pass through the origin ($%
J=K=0$). Remarkably, the 5K points lie reasonably close to the dashed line,
giving us hope that the expression (\ref{J-Kmin}) might be quite good.
Before detailing quantitative aspects of the analysis, let us comment on a
remarkable aspect of this data. From the 5K simulated $J$'s, we compiled
histograms to form a current distribution for each $\nu $. 

%================================
\begin{figure}[tbp]
\begin{center}
\includegraphics[height=8cm,width=10cm]{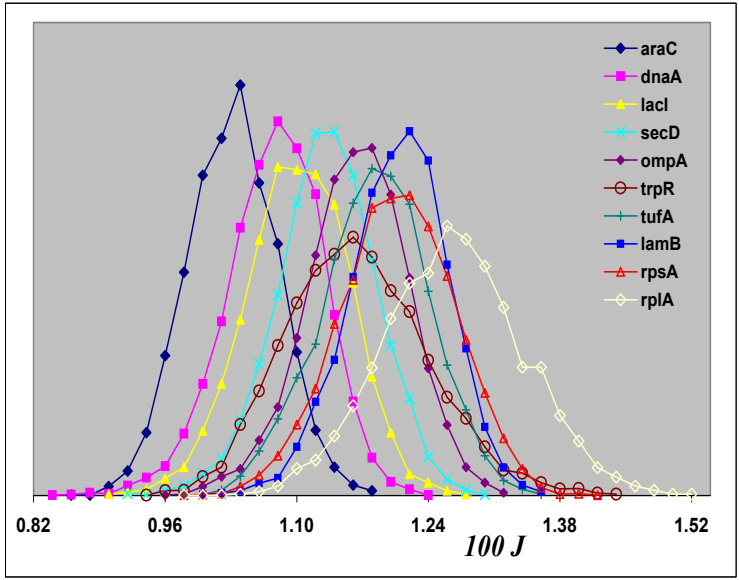}
\caption{Current distributions for 5000 TASEP sequences (modeling the 
silent mutations) which code for the 10 indicated proteins. }
\label{fig:F7}
\end{center}
\end{figure}
%================================================

Shown in Fig.~\ref{fig:F7}, 
no major surprises are apparent: All distributions seem normal, with means
in the approximate range of $1.00$-$1.25$ (for $100\times J$) and standard
deviations of $0.05$-$0.10$. Their skewness and kurtosis (both unitless
measures of deviations from pure Gaussians) fall in the ranges of,
respectively, $\left[ -0.3,0.3\right] $ and $\left[ -0.1,0.4\right] $.
However, with even a casual glance at the panels in Fig.~6, the reader may
notice that all but two of the wild types lie well {\em above} the cluster
of 5K points. Indeed, five of them are more than 6.5 standard deviations
above the mean. We can only speculate that natural evolution optimized the
production rates! Work is in progress to study the rest of the 5416 proteins
in {\em E.coli} and, if this systematic deviation persists, to consider
possible deeper underpinnings of this phenomenon.

Returning to the more practical issue at hand, we seek a quantitative
description in an attempt to test expression (\ref{J-Kmin}). For a
particular protein $\nu $, we consider an ensemble in which all $M_{\nu }$
sequences $\left\{ c_{i}\right\} _{\nu }$ are equally probable. However, as
illustrated by the last two rows of table (\ref{aa-deg}), each distinct 
$\left\{ \gamma _{i}\right\} _{\nu }$ sequence can result from several 
$\left\{ c_{i}\right\} _{\nu }$ sequences (depending on conditions on the
cell, and other complications which we ignore here). Due to this complicated 
$c_{i}$-$\gamma _{i}$ connection, there are far fewer $\left\{ \gamma
_{i}\right\} _{\nu }$ sequences, so that the distribution of $\gamma $'s,
which we denote by ${\cal P}_{\nu }\left[ \left\{ \gamma _{i}\right\}
\right] $, will not be trivial. Since $J$ depends only on $\left\{ \gamma
_{i}\right\} $ and not $\left\{ c_{i}\right\} $, this ${\cal P}_{\nu }$ will
control the average $J$ over our ensemble. Of course, ${\cal P}_{\nu }\left[
\left\{ \gamma _{i}\right\} \right] $ is still a product distribution, $%
\prod_{i=1}^{L_{\nu }}p_{i}\left( \gamma _{i}\right) $, since no
correlations between sites are assumed. But, unlike previous studies, the $%
\gamma $'s here must be chosen from {\em site-dependent} distributions --
thus the subscript $i$ on $p_{i}$. To clarify, let us return to our
illustration, in which ${\cal P}_{\mbox{IWAMS}}\left[ \left\{ \gamma
_{i}\right\} \right] =\prod_{i=1}^{5}p_{i}\left( \gamma _{i}\right) $ with,
explicitly, 
\begin{eqnarray*}
p_1\left( \gamma \right)  &=&\delta \left( \gamma -0.421\right)  \\
p_2\left( \gamma \right)  &=&\delta \left( \gamma -0.171\right)  \\
p_3\left( \gamma \right)  &=&\left. \left\{ \delta \left( \gamma
-0.122\right) +3\delta \left( \gamma -0.695\right) \right\} \right/ 4 \\
p_4\left( \gamma \right)  &=&\delta \left( \gamma -0.477\right)  \\
p_5\left( \gamma \right)  &=&\left\{ \delta \left( \gamma -0.137\right)
+2\delta \left( \gamma -0.193\right) +\delta \left( \gamma -0.251\right)
\right.  \\
&&\left. \left. \,\,+\delta \left( \gamma -0.300\right) +\delta \left(
\gamma -0.388\right) \right\} \right/ 6
\end{eqnarray*}
Since the sequence of $p_{i}$'s are fixed by the amino acid sequence, $%
\left\{ aa_{i}\right\} $, we are constrained by a quenched distribution of
different $p$'s. Thus, we arrive at the notion of a quenched ``distribution
of distributions.''

With this framework in mind, we can define another average of $J$,
associated with all possible ways of producing protein $\nu $ in our
ensemble of $M_{\nu }$ (equally probable) silent mutations: 
\[
<J>_{\nu }=\int {\cal D}\gamma \,\,J\left( \left\{ \gamma _{i}\right\}
\right) {\cal P}_{\nu }\left[ \left\{ \gamma _{i}\right\} \right] 
\]
where ${\cal D}\gamma $ denotes $\prod_{i=1}^{L_{\nu }}d\gamma _{i}$. In a
similar vein, $K_{12,\min }$ also depends only on $\left\{ \gamma
_{i}\right\} $ (and not $\left\{ c_{i}\right\} $), so that 
\[
<K_{12,\min }>_{\nu }=\int {\cal D}\gamma \,\,K_{12,\min }\left( \left\{
\gamma _{i}\right\} \right) {\cal P}_{\nu }\left[ \left\{ \gamma
_{i}\right\} \right] 
\]
The simulations for each $\nu $ in Fig.~6 are a 5K-point sampling of this 
${\cal P}_{\nu }$. Thus, the coordinates of the ``center of mass'' of the
(roughly elliptical) cluster are just $<J>_{\nu }$ and $<K_{12,\min }>_{\nu
} $. Of course, we can consider other quantities of interest, such as 
$<\delta \left( J-J\left( \left\{ \gamma _{i}\right\} \right) \right) >_{\nu
} $, corresponding to the histograms in Fig.~7. Other obvious possibilities
are the second moments, which will provide us with the two axes and the
orientation of each cluster, as well as a measure of the $J$-$K$
correlation. Here, we are content to focus only the averages and their
ratios, $<J>_{\nu }/<K_{12,\min }>_{\nu }$, for these 10 proteins.
Remarkably, though both $<J>_{\nu }$ and $<K>_{\nu }$ range by 25\% (over
the 10 $\nu $'s), this ratio is essentially constant! This observation
motivates us to define $A$ in (\ref{J-Kmin}) by a further average:

\[
A\equiv \frac{1}{10}\sum_{\nu }\frac{<J>_{\nu }}{<K_{12,\min }>_{\nu }}\,\,. 
\]

From our data, we find 
\[
A\cong 0.0656 
\]
which is, interestingly, comparable to $\left( 1+\sqrt{12}\right) ^{-2}\cong
0.0502$. As a test of its ``predictive power,'' we computed $AK_{12,\min }$
for all $5000\times 10$ $\left\{ \gamma \right\} $'s and compared them to
the values of the currents obtained from simulations. Specifically, the
average of (these 50K values of) $AK_{12,\min }/J$ is within $0.4\%$ of
unity, while the standard deviation is about $5\%$. Rarely does this ratio
range more than 15\% from unity. In this sense, we are hopeful that, when we
extend this study to the other 5406 genes in {\em E.coli}, we will confirm $%
AK_{12,\min }\left( \left\{ \gamma _{i}\right\} \right) $ as a simple and
reliable estimate for $J\left( \left\{ \gamma _{i}\right\} \right) $.

Despite being quite involved and extensive, this study has answered few
questions in biology. Though it remains far from the goal of understanding
protein synthesis in real cells, it does pose rich new ground for exploring
nonequilibrium statistical mechanics. The main progress here is that,
should we wish to design silent mutations of genes that could outperform the
wild type, either by enhancing or suppressing the production rates of this
protein, a reliable and simple method is available to facilitate the search
for our goals.

\section{\label{sum}Conclusions and outlook}

In this article, we have touched upon two fundamental issues in two very
different fields:\ understanding nonequilibrium steady states and
developing quantitative models for protein production. These two seemingly
disparate problems converge in a simple one-dimensional transport model, the
totally asymmetric simple exclusion process and its modifications. The
TASEP\ is a paradigmatic far-from-equilibrium model, characterized by open
boundaries and a systematic particle current through the system. Due to the
exclusion, the particles are interacting, and so it is highly non-trivial to
find steady-state and dynamic properties. Still, a considerable body of
exact results is available for the standard TASEP. In particular, despite
the one-dimensional nature of the model, it displays three distinct phases,
separated by first order and continuous transitions. If the model is
modified to include extended particles and inhomogeneous hopping rates, it
is generally accessible only via simulations or approximate (mean-field)
methods.

With these modifications, the model becomes a more realistic - but still
highly simplified - description of protein synthesis. The one-dimensional
lattice models the mRNA template, with sites and extended particles
representing codons and ribosomes, respectively. Further, we allow
non-uniform hopping rates, to reflect the variability of the aa-tRNA
concentrations associated with different codons. The particle current
through the TASEP is simply the protein production rate. An interesting
feature of translation is the sophisticated degeneracy: 61 codons code for
20 amino acids (mediated by 46 tRNAs in {\em E.coli}). In other words, there
are many distinct sequences (``silent mutations'') which code for the same
protein but are characterized by different production rates.

In this article, we presented a brief introduction to the main findings for
TASEP and the basics of protein synthesis, designed with non-experts in
mind. We also described the modeling of translation in terms of a
generalized TASEP, summarizing both well-established and more recent
results. Amongst the latter we discussed two specific topics: first, the
effects of limited availability of particles, and second, simple but
remarkably good estimates for currents in the fully inhomogeneous case. The
first project is motivated by the observation that ribosomes are large
molecules so that their synthesis is costly for the cell. Hence, it is
reasonable to expect them to be in limited supply. Considering only the
simplest case - fully uniform rates and particles covering only one site -
we asked:\ How are currents and density profiles affected if a single, or
several, TASEPs compete for particles from a finite reservoir? Remarkable
results, such as multiple, distinct regimes in density profiles and shock
localization were discovered. The second discussion centered on two
questions: Is it possible to arrive at simple yet reliable estimates for
currents associated with fully inhomogeneous sequences? And how do the
currents associated with the ``ensemble'' of silent mutations compare to
that of the wild type? The answer to the first question relies on computing
the typical time any particular codon is covered by a ribosome. In the
language of TASEP, this is the time it takes a particle to traverse a
stretch of $12$ sites around a given site. This quantity can be determined
from sequence information with minimal effort (provided the aa-tRNA
concentrations are known). Its inverse is effectively a coarse-grained rate
associated with hopping from that site. It turns out that the lowest of
these rates (in a given sequence), denoted by $K_{12,\min }$, provides a
good estimate for the average current. Specifically, Monte Carlo simulations
for $5000$ randomly selected silent mutations of $10$ different proteins
show a reliable linear relation between currents and $K_{12,\min }$'s, with
a proportionality constant that appears to be the {\em same} for all the
proteins studied. Moreover, the current and $K_{12,\min }$ of the wild type
also obey this linear relation even though both fall well above the typical
values for randomly chosen sequences.

Clearly, the explorations reported here leave many questions unanswered,
both on the statistical physics and the biology side. We just cite a few
which will hopefully spark future research. The central fundamental question
concerns the ``stability'' of steady-state properties with respect to model
modifications. Which changes of microscopic model details (e.g., hopping
rates)\ will lead to changes of microscopic or macroscopic behaviors?\ While
notions of universality and independence from certain dynamic details are
well understood for equilibrium systems, we have taken only initial steps
towards extending them to nonequilibrium steady states \cite{ZS06,ZS07}.
Further, little if anything is known about how these general concepts apply
to specific models. On the quantitative biology side, even relatively simple
questions remain open: Are aa-tRNA concentrations really the limiting
factor for protein production rates? Are there other intrinsic rates, or is
initiation the critical bottle neck? Secondary structures are known to be
important \cite{KMTP09}, but how exactly do they affect production rates?\
Why are the currents of wild type genes so optimized? Clearly, fundamental
insights and close collaborations between physicists and biologists are
needed before we will begin to understand biological processes -- which are
generically far from equilibrium -- at a quantitative level.

\begin{acknowledgements}
We are grateful for numerous illuminating discussions with many colleagues, 
especially T. Chou, L.J. Cook, C.V. Finkielstein, R.J. Harris, K. Mallick, 
and L.B. Shaw. We thank J.L. Lebowitz for the invitation to participate in 
this special issue of the Journal of Statistical Physics. This work is 
supported in part by grants from the NSF: DMR-0705152 and DMR-1005417.
\end{acknowledgements}

 \bibliographystyle{plain} 
\bibliography{ZDS_ref}

\begin{thebibliography}{100}

\bibitem{ASZ08}
D.A. Adams, B~Schmittmann, and R.~K.~P. Zia.
\newblock Far-from-equilibrium transport with constrained resources.
\newblock {\em Journal of Statistical Mechanics: Theory and Experiment},
  2009:P06009, 2008.
\newblock For a similar TASEP with fixed total particle number, see
  \cite{HdN02}.

\bibitem{AZS07}
D.A. Adams, R.~K.~P. Zia, and B~Schmittmann.
\newblock Power spectra of the total occupancy in the totally asymmetric simple
  exclusion process.
\newblock {\em Physical Review Letters}, 99:020601, 2007.

\bibitem{cell}
B.~Alberts, A.~Johnson, J.~Lewis, M.~Raff, K.~Roberts, and P.~Walter.
\newblock {\em How Cells Read the Genome: From DNA to Protein}.
\newblock Garland Science, 2007.

\bibitem{AL03}
F.~C. Alcaraz and M.~J. Lazo.
\newblock The exact solution of the asymmetric exclusion problem with particles
  of arbitrary size: Matrix product ansatz.
\newblock {\em Brazilian Journal of Physics}, 33:533, 2003.

\bibitem{APLE06}
A.~Antoun, M.~Y. Pavlov, M.~Lovmar, and M.~Ehrenberg.
\newblock How initiation factors tune the rate of initiation of protein
  synthesis in bacteria.
\newblock {\em EMBO J}, 25(11):2539--2550, 2006.

\bibitem{BCCCCOPPVZ08}
M.~Ballerini, N.~Cabibbo, R.~Candelier, A.~Cavagna, E.~Cisbani, I.~Giardina,
  A.~Orlandi, P.~Parisi, A.~Procaccini, M.~Viale, and V.~Zdravkovic.
\newblock Empirical investigation of starling flocks: a benchmark study in
  collective animal behaviour.
\newblock {\em Animal Behaviour}, 76:201, 2008.

\bibitem{BS02}
V.~Belitzky and G.~M. Sch\"{u}tz.
\newblock Diffusion and scattering of shocks in the partially asymmetric
  exclusion process.
\newblock {\em Electron. J. Probab.}, 7(11):1, 2002.
\newblock Also see \cite{KSKS98}.

\bibitem{BBEM99}
M.~Bengrine, A.~Benyoussef, H.~Ez-Zahraouy, and F.~Mhirech.
\newblock Traffic model with quenched disorder.
\newblock {\em Phys. Lett. A}, 253(3-4):135, 1999.

\bibitem{BSWN94}
D.~Beyer, E.~Skripkin, J.~Wadzack, and K.~H. Nierhaus.
\newblock How the ribosome moves along the mrna during protein synthesis.
\newblock {\em Journal of biological chemistry}, 269(48):30713--30717, 1994.

\bibitem{proteinsyn}
BioStudio.
\newblock \texttt{www.biostudio.com/demo\_freeman\_protein\_synthesis.htm}.

\bibitem{BE07}
R.~A. Blythe and M.~R. Evans.
\newblock Nonequilibrium steady states of matrix-product form: a solver's
  guide.
\newblock {\em J. Phys. A: Math. Gen.}, 40(46):R333, 2007.

\bibitem{Chou03}
T.~Chou.
\newblock Ribosome recycling, diffusion, and mrna loop formation in
  translational regulation.
\newblock {\em Biophysical Journal}, 85:755, 2010.

\bibitem{LC03}
T.~Chou and G.~Lakatos.
\newblock Totally asymmetric exclusion processes with particles of arbitrary
  size.
\newblock {\em J. Phys. A: Math. Gen.}, 36:2027, 2003.

\bibitem{CL04}
T.~Chou and G.~Lakatos.
\newblock Clustered bottlenecks in mrna translation and protein synthesis.
\newblock {\em Phys. Rev. Lett.}, 93:198101, 2004.

\bibitem{CSS00}
D.~Chowdhury, L.~Santen, and A.~Schadschneider.
\newblock Statistical physics of vehicular traffic and some related systems.
\newblock {\em Phys. Rep.}, 329:199, 2000.

\bibitem{CSS99}
D.~Chowdhury, L.~Santen, and A.~Schadschneider.
\newblock Vehicular traffic: A system of interacting particles driven far from
  equilibrium.
\newblock {\em Curr. Sci.}, 77:411, 2000.

\bibitem{DOE}
Basic Energy Sciences~Advisory Committee.
\newblock {\em Directing Matter and Energy: Five Challenges for Science and the
  Imagination}.
\newblock Washington, DC: Department of Energy Publications, 2007.

\bibitem{Cook10}
L.~J. Cook.
\newblock private communication, 2010.

\bibitem{CZ09}
L.J. Cook and R.~K.~P. Zia.
\newblock Feedback and fluctuations in a totally asymmetric simple exclusion
  process with finite resources.
\newblock {\em Journal of Statistical Mechanics: Theory and Experiment},
  2009:P02012, 2009.

\bibitem{CZ10}
L.J. Cook and R.~K.~P. Zia.
\newblock Power spectra of a constrained totally asymmetric simple exclusion
  process.
\newblock {\em Journal of Statistical Mechanics: Theory and Experiment},
  2010:P07014, 2010.

\bibitem{CZS09}
L.J. Cook, R.~K.~P. Zia, and B~Schmittmann.
\newblock Competition between many totally asymmetric simple exclusion
  processes for a finite pool of resources.
\newblock {\em Physical Review E}, 80(3):031142, 2009.

\bibitem{Crick58}
F.H.C. Crick.
\newblock On protein synthesis.
\newblock {\em Symp. Soc. Exp. Biol.}, XII:139 -- 163, 1958.

\bibitem{Crick70}
F.H.C. Crick.
\newblock Central dogma of molecular biology.
\newblock {\em Nature}, 227:561 -- 563, 1970.

\bibitem{CM96}
J.~Czworkowski and P.~B. Moore.
\newblock The elongation phase of protein synthesis.
\newblock {\em Progress in nucleic acid research and molecular biology},
  54:293--332, 1996.

\bibitem{dGE06}
J.~de~Gier and F.~H.~L. Essler.
\newblock Exact spectral gaps of the asymmetric exclusion process with open
  boundaries.
\newblock {\em Journal of Statistical Mechanics: Theory and Experiment},
  2006:P12011, 2006.

\bibitem{dMF85}
A.~De~Masi and P.~A. Ferrari.
\newblock Self diffusion in one dimensional lattice gases in the presence of an
  external field.
\newblock {\em J. Stat. Phys.}, 38:603, 1985.

\bibitem{Derrida07}
B.~Derrida.
\newblock Non-equilibrium steady states: fluctuations and large deviations of
  the density and of the current.
\newblock {\em J. Stat. Mech.}, 2007:P07023, 2007.

\bibitem{DDM92}
B.~Derrida, E.~Domany, and D.~Mukamel.
\newblock An exact solution of a one-dimensional asymmetric exclusion model
  with open boundaries,.
\newblock {\em J. Stat.Phys.}, 69:667, 1992.

\bibitem{DEP93}
B.~Derrida, M.~R. Evans, and V.~Pasquier.
\newblock Exact solution of a 1d asymmetric exclusion model using a matrix
  formulation.
\newblock {\em J. Phys. A: Math. Gen.}, 26:1493, 1993.

\bibitem{DEHP93}
B.~Derrida, M.R. Evans, V.~Hakim, and V.~Pasquier.
\newblock Exact solution of a 1d asymmetric exclusion model using a matrix
  formulation.
\newblock {\em J. Phys. A: Math. Gen.}, 26:1493, 1993.

\bibitem{DG09}
B.~Derrida and A.~Gerschenfeld.
\newblock Current fluctuations of the one dimensional symmetric simple
  exclusion process with step initial condition.
\newblock {\em J. Stat. Phys.}, 136(1):1, 2009.

\bibitem{DJLS93}
B.~Derrida, S.~A. Janowsky, J.~L. Lebowitz, and E.~R. Speer.
\newblock Exact solution of the totally asymmetric simple exclusion process:
  shock profiles.
\newblock {\em J. Stat. Phys}, 73:813, 1993.

\bibitem{DNK96}
H.~Dong, L.~Nilsson, and C.G. Kurland.
\newblock Co-variation of trna abundance and codon usage in escherichia coli at
  different growth rates.
\newblock {\em Journal of Molecular Biology}, 260(5):649--63, 1996.

\bibitem{DSZ07_PRE}
J.J. Dong, B~Schmittmann, and R.~K.~P. Zia.
\newblock Inhomogeneous exclusion processes and protein synthesis.
\newblock {\em Physical Review E}, 76:051113, 2007.

\bibitem{DSZ07_JSP}
J.J. Dong, B~Schmittmann, and R.~K.~P. Zia.
\newblock Towards a model for protein production rates.
\newblock {\em Journal of Statistical Physics}, 128(1):21, 2007.

\bibitem{DZS09}
J.J. Dong, R.~K.~P. Zia, and B~Schmittmann.
\newblock Understanding the edge effect in tasep with mean-field theoretic
  approaches.
\newblock {\em Journal of Physics A: Mathematical and Theoretical},
  42(1):015002, 2009.

\bibitem{Dotsenko01}
V.~Dotsenko.
\newblock {\em Introduction to the Replica Theory of Disorderred Statistical
  Systems}.
\newblock Cambridge University Press, 2001.

\bibitem{DS00}
M.~Dudzinski and G.~M. Sch\"{u}tz.
\newblock Relaxation spectrum of the asymmetric exclusion process with open
  boundaries.
\newblock {\em J. Phys. A: Math. Gen.}, 33(47):8351, 2000.

\bibitem{EA75}
S.~F. Edwards and P.~W. Anderson.
\newblock Theory of spin glasses.
\newblock {\em J. Phys. F: Metal Physics}, 5:965, 1975.

\bibitem{Feder07}
T.~Feder.
\newblock Statistical physics is for the birds.
\newblock {\em Physics Today}, 60(10):28, 2007.

\bibitem{FKT08}
M.~Ebrahim Foulaadvand, Anatoly~B. Kolomeisky, and H.~Teymouri.
\newblock Asymmetric exclusion processes with disorder: Effect of correlations.
\newblock {\em Phys. Rev. E}, 78(6):061116, Dec 2008.

\bibitem{FH05}
C.~S. Fraser and J.~W.~B. Hershey.
\newblock Movement in ribosome translocation.
\newblock {\em Journal of Biology}, 4(2):8, 2005.

\bibitem{Friedman08}
M.~H. Friedman.
\newblock {\em Principles and Models of Biological Transport}.
\newblock Berlin: Springer, 2008.

\bibitem{GM04}
O.~Golinelli and K.~Mallick.
\newblock Bethe ansatz calculation of the spectral gap of the asymmetric
  exclusion process.
\newblock {\em J. Phys. A: Math. Gen.}, 37(10):3321, 2004.

\bibitem{GM05}
O.~Golinelli and K.~Mallick.
\newblock Spectral gap of the totally asymmetric exclusion process at arbitrary
  filling.
\newblock {\em J. Phys. A: Math. Gen.}, 38(7):1419, 2005.
\newblock For recent reviews, see e.g., \cite{GM06,BE07,Derrida07}.

\bibitem{GM06}
O.~Golinelli and K.~Mallick.
\newblock The asymmetric simple exclusion process: an integrable model for
  non-equilibrium statistical mechanics.
\newblock {\em J. Phys. A: Math. Gen.}, 39(41):12679, 2006.

\bibitem{GS08}
P.~Greulich and A~Schadschneider.
\newblock Single-bottleneck approximation for driven lattice gases with
  disorder and open boundary conditions.
\newblock {\em Journal of Statistical Mechanics: Theory and Experiment},
  2008(P04009), 2008.

\bibitem{GMGB07}
S.~Gupta, S.~N. Majumdar, C.~Godr\`{e}che, and M.~Barma.
\newblock Tagged particle correlations in the asymmetric simple exclusion
  process: Finite-size effects.
\newblock {\em Phys. Rev. E}, 76:021112, 2007.

\bibitem{GS92}
L-H Gwa and H.~Spohn.
\newblock Bethe solution for the dynamical scaling exponent of the noisy
  burgers equation.
\newblock {\em Phys. Rev. A}, 46:844, 1992.

\bibitem{HdN02}
M.~Ha and M~den Nijs.
\newblock Macroscopic car condensation in a parking garage.
\newblock {\em Phys. Rev. E}, 66:036118, 2002.

\bibitem{HS04}
R.~J. Harris and R.~B. Stinchcombe.
\newblock Disordered asymmetric simple exclusion process: Mean-field treatment.
\newblock {\em Phys. Rev. E}, 70:016108, 2004.

\bibitem{HR80}
R.~Heinrich and T.~A. Rapoport.
\newblock Mathematical modelling of translation of mrna in eucaryotes; steady
  states, time-dependent processes and application to reticulocytest.
\newblock {\em Journal of Theoretical Biology}, 86(2):279 -- 313, 1980.

\bibitem{Howard01}
J.~Howard.
\newblock {\em Mechanics of Motor Proteins and the Cytoskeleton}.
\newblock Sunderland, MA: Sinauer, 2001.

\bibitem{Ikemura81}
T.~Ikemura.
\newblock Correlation between the abundance of escherichia coli transfer rnas
  and the occurrence of the respective codons in its protein genes.
\newblock {\em Journal of molecular biology}, 146(1):1--21, 1981.

\bibitem{JL92}
S.~A. Janowsky and J.~L. Lebowitz.
\newblock Finite-size effects and shock fluctuations in the asymmetric
  simple-exclusion process.
\newblock {\em Phys. Rev. A}, 45:618, 1992.

\bibitem{Kim95}
D.~Kim.
\newblock Bethe ansatz solution for crossover scaling functions of the
  asymmetric xxz chain and the kardar-parisi-zhang-type growth model.
\newblock {\em Phys. Rev. E}, 52:3512, 1995.

\bibitem{KF99}
L.L. Kisselev and L.~Y. Frolova.
\newblock Termination of translation in eukaryotes: new results and new
  hypotheses.
\newblock {\em Biochemistry (Moscow)}, 64:8--16, 1999.

\bibitem{KF07}
A.~B. Kolomeisky and M.~E. Fisher.
\newblock Molecular motors:a theorist's perspective.
\newblock {\em Annu. Rev. Phys. Chem.}, 58:675, 2007.

\bibitem{KSKS98}
A.~B. Kolomeisky, G.M. Sch\"{u}tz, E.B. Kolomeisky, and J.P. Straley.
\newblock Phase diagram of one-dimensional driven lattice gases with open
  boundaries.
\newblock {\em J. Phys. A: Math. Gen.}, 31(33):6911, 1998.

\bibitem{Kolomeisky98}
Anatoly~B. Kolomeisky.
\newblock Asymmetric simple exclusion model with local inhomogeneity.
\newblock {\em J. Phys. A: Math. Gen.}, 31(4):1153, Jan 1998.

\bibitem{Kozak83}
M.~Kozak.
\newblock Comparison of initiation of protein synthesis in procaryotes,
  eucaryotes, and organelles.
\newblock {\em Microbological Reviews}, 47(1):1--45, 1983.

\bibitem{Krug91}
J.~Krug.
\newblock Boundary-induced phase transitions in driven diffusive systems.
\newblock {\em Phys. Rev. Lett.}, 67:1882, 1991.

\bibitem{Krug00}
J.~Krug.
\newblock Phase separation in disordered exclusion models.
\newblock {\em BJP}, 30:97--104, 2000.

\bibitem{KMTP09}
G.~Kudla, A.~W. Murray, D.~Tollervey, and J.~B. Plotkin.
\newblock Coding-sequence determinants of gene expression in escherichia coli.
\newblock {\em Science}, 324(5924):255--258, 2009.

\bibitem{KvB85}
R.~Kutner and H.~van Beijeren.
\newblock Influence of an external force on tracer diffusion in a
  one-dimensional lattice gas.
\newblock {\em J. Stat. Phys.}, 39:317, 1985.

\bibitem{LSMS05}
B.~S. Laursen, H.~P. S{\o}rensen, K.~K. Mortensen, and H.~U. Sperling-Petersen.
\newblock Initiation of protein synthesis in bacteria.
\newblock {\em Microbiology and molecular biology reviews}, 69(1):101, 2005.

\bibitem{Maaloe79}
O.~Maaloe.
\newblock Regulation of the protein-synthesizing machinery - ribosomes, trna,
  factors and so on.
\newblock In R.F. Goldberger, editor, {\em Biological Regulation and
  Development}. New York: Plenum Press, 1979.

\bibitem{MG69}
C.~T. MacDonald and J.~H. Gibbs.
\newblock Concerning the kinetics of polypeptide synthesis on polyribosomes.
\newblock {\em Biopolymers}, 7:707, 1969.

\bibitem{MGP68}
C.~T. MacDonald, J.~H. Gibbs, and A.~C. Pipkin.
\newblock Kinetics of biopolymerization on nucleic acid templates.
\newblock {\em Biopolymers}, 6(1), 1968.

\bibitem{MB91}
S.~N. Majumdar and M.~Barma.
\newblock Tag diffusion in driven systems, growing interfaces, and anomalous
  fluctuations.
\newblock {\em Phys. Rev. B}, 44:5306, 1991.

\bibitem{Mallick96}
K.~Mallick.
\newblock Shocks in the asymmetry exclusion model with an impurity.
\newblock {\em J. Phys. A: Math. Gen.}, 29:5375, 1996.

\bibitem{Merrick92}
W.~Merrick.
\newblock Mechanism and regulation of eukaryotic protein synthesis.
\newblock {\em Microbiology and molecular biology reviews}, 56(2):291, 1992.

\bibitem{Moldave85}
K.~Moldave.
\newblock Eukaryotic protein synthesis.
\newblock {\em Annual review of biochemistry}, 54(1):1109--1149, 1985.

\bibitem{NAS02}
Z.~Nagy, C.~Appert, and L.~Santen.
\newblock Relaxation times in the asep model using a dmrg method.
\newblock {\em J. Stat. Phys.}, 109:623, 2002.

\bibitem{NI98}
Y.~Nakamura and K.~Ito.
\newblock How protein reads the stop codon and terminates translation.
\newblock {\em Genes to Cells}, 3:265--278, 1998.

\bibitem{NE98}
B.~Negrutskii and A.~El'Skaya.
\newblock Eukaryotic translation elongation factor 1 [alpha]: Structure,
  expression, functions, and possible role in aminoacyl-trna channeling.
\newblock {\em Progress in nucleic acid research}, 60:47--78, 1998.

\bibitem{NN90}
O.~Nyg{\aa}rd and L.~Nilsson.
\newblock Translational dynamics: Interactions between the translational
  factors, trna and ribosomes during eukaryotic protein synthesis.
\newblock {\em European Journal of Biochemistry}, 191(1):1--17, 1990.

\bibitem{NAS}
Board on~Physics and Astronomy.
\newblock {\em Condensed-Matter and Materials Physics: The Science of the World
  Around Us}.
\newblock Washington, DC: The National Academies Press, 2007.

\bibitem{PH00}
T.~Pestova and C.~Hellen.
\newblock The structure and function of initiation factors in eukaryotic
  protein synthesis.
\newblock {\em Cellular and Molecular Life Sciences}, 57(4):651, 2000.

\bibitem{PPOF05}
P.~Pierobon, A.~Parmeggiani, F.~von Oppen, and E.~Frey.
\newblock Dynamic correlation functions and boltzmann langevin approach for
  driven one dimensional lattice gas.
\newblock {\em Phys. Rev. E}, 72:036123, 2005.

\bibitem{PSSS01}
V.~Popkov, L.~Santen, A.~Schadschneider, and G.~M. Sch\"{u}tz.
\newblock Boundary-induced phase transitions in traffic flow.
\newblock {\em J. Phys. A: Math. Gen.}, 34:L45, 2001.

\bibitem{RRCM90}
B.~Riis, S.~Rattan, B.~Clark, and W.~Merrick.
\newblock Eukaryotic protein elongation factors.
\newblock {\em Trends in biochemical science}, 15:420--424, 1990.

\bibitem{SA02}
L.~Santen and C.~Appert.
\newblock The asymmetric exclusion process revisited: Fluctuations and dynamics
  in the domain wall picture.
\newblock {\em J. Stat. Phys.}, 106:187, 2002.

\bibitem{SW03}
M.~Schliwa and G.~Woehlke.
\newblock Molecular motors.
\newblock {\em Nature}, 422:759, 2003.

\bibitem{Schutz93}
G.M. Sch\"{u}tz.
\newblock Time-dependent correlation functions in a one-dimensional asymmetric
  exclusion process.
\newblock {\em Phys. Rev. E}, 47:4265, 1993.

\bibitem{SD93}
G.M. Sch\"{u}tz and E.~Domany.
\newblock Phase transitions in an exactly solvable one-dimensional exclusion
  process.
\newblock {\em J. Stat. Phys.}, 72:277, 1993.

\bibitem{SKL04}
L.~B. Shaw, A.~B. Kolomeisky, and K.~H. Lee.
\newblock Local inhomogeneity in asymmetric simple exclusion processes with
  extended objects.
\newblock {\em J. Phys. A: Math. Gen.}, 37:2105, 2004.

\bibitem{SZL03}
L.B. Shaw, R.~K.~P. Zia, and K.H. Lee.
\newblock Modeling, simulations, and analyses of protein synthesis: Driven
  lattice gas with extended objects.
\newblock {\em Physical Review E}, 68:021910, 2003.

\bibitem{SD75}
J.~Shine and L.~Dalgarno.
\newblock Determinant of cistron specificity in bacterial ribosomes.
\newblock {\em Nature}, 254(5495):34, 1975.

\bibitem{SP91}
M.~A. S{\o}rensen and S.~Pedersen.
\newblock Absolute in vivo translation rates of individual codons in
  escherichia coli : The two glutamic acid codons gaa and gag are translated
  with a threefold difference in rate.
\newblock {\em Journal of Molecular Biology}, 222(2):265 -- 280, 1991.

\bibitem{SKPF03}
L.~Spector, J.~Klein, C.~Perry, and M.~Feinstein.
\newblock Emergence of collective behavior in evolving populations of flying
  agents.
\newblock In E.~Cantu-Paz, J.~A. Foster, K.~Deb, L.~D. Davis, R.~Roy, U-M
  O'Reilly, H-G Beyer, R.~Standish, G.~Kendall, S.~Wilson, M.~Harman,
  J.~Wegener, D.~Dasgupta, M.~A. Potter, A.~C. Schultz, K.~A. Dowsland,
  N.~Jonoska, and J.~Miller, editors, {\em Proceedings of the Genetic and
  Evolutionary Computation Conference}. Berlin: Springer, 2003.

\bibitem{Spitzer70}
F~Spitzer.
\newblock Interaction of markov processes.
\newblock {\em Adv. Math.}, 5:246, 1970.

\bibitem{TMH03}
S.~Takesue, T.~Mitsudo, and H.~Hayakawa.
\newblock Power-law behavior in the power spectrum induced by brownian motion
  of a domain wall.
\newblock {\em Phys. Rev. E}, 68:015103(R), 2003.

\bibitem{TB98}
G.~Tripathy and M.~Barma.
\newblock Driven lattice gases with quenched disorder: Exact results and
  different macroscopic regimes.
\newblock {\em Phys. Rev. E}, 58(1911), 1998.

\bibitem{VBL84}
S.~Varenne, J.~Buc, R.~Lloubes, and C.~Lazdunski.
\newblock Translation is a non-uniform process: Effect of trna availability on
  the rate of elongation of nascent polypeptide chains.
\newblock {\em Journal of molecular biology}, 180(3):549--576, 1984.

\bibitem{genecode}
Wikipedia.
\newblock Genetic code.
\newblock \texttt{en.wikipedia.org/wiki/Genetic\_code}.

\bibitem{YY98}
A.~P. Young, editor.
\newblock {\em Spin Glasses and Random Fields}.
\newblock World Scientific, 1998.

\bibitem{APS09}
R.~K.~P. Zia, J.~J. Dong, and B.~Schmittmann.
\newblock Estimating currents in totally asymmetric simple exclusion process
  with extended particles and inhomogeneous hopping rates.
\newblock \texttt{meetings.aps.org/Meeting/MAR09/Event/93724}.

\bibitem{ZS06}
R.~K.~P. Zia and B~. Schmittmann.
\newblock A possible classification of nonequilibrium steady states.
\newblock {\em J. Phys. A: Math. Gen.}, 39(41):L407 -- L413, 2006.

\bibitem{ZS07}
R.~K.~P. Zia and B~. Schmittmann.
\newblock Probability currents as principal characteristics in the statistical
  mechanics of nonequilibrium steady states.
\newblock {\em Journal of Statistical Mechanics: Theory and Experiment},
  2007:P07012, 2007.

\end{thebibliography}

\end{document}